\documentclass[aoas]{imsart}

\RequirePackage{amsthm,amsmath,amsfonts,amssymb}
\RequirePackage[authoryear]{natbib}
\RequirePackage[colorlinks,citecolor=blue,urlcolor=blue]{hyperref} % ok with imsart if loaded early
\RequirePackage{graphicx}

\usepackage{bm}
\usepackage{adjustbox}
\usepackage[table]{xcolor}
\usepackage{enumitem}
\usepackage{makecell}
\usepackage{subcaption}       
\usepackage{bbold}
\usepackage{mathtools}
\usepackage{colortbl}
\usepackage{longtable}
\usepackage{comment}
\usepackage{microtype}
\usepackage{booktabs}
\usepackage{algorithm}
\usepackage{algorithmic}
\usepackage{listings}
\usepackage{multicol}
\usepackage{xparse}

\usepackage{tikz}
\usetikzlibrary{shapes.geometric,fit,positioning,quotes,calc}

\definecolor{color1}{rgb}{0.83,0.79,0.85}
\definecolor{color2}{rgb}{0.67,0.62,0.71}
\definecolor{wimbledonGreen}{HTML}{2E8B57}
\definecolor{clayOrange}{HTML}{D2691E}
\definecolor{hardcourtBlue}{HTML}{1E90FF}
\definecolor{color3}{rgb}{0.60,0.76,0.75}
\hypersetup{
  bookmarksopen=true,
  bookmarksdepth=3
}
\startlocaldefs

\endlocaldefs

\endlocaldefs

\begin{document}

\begin{frontmatter}
\title{The Bradley--Terry Stochastic Block Model}
%\title{A sample article title with some additional note\thanksref{T1}}
\runtitle{BT--SBM}

\begin{aug}

\author[A]{\fnms{Lapo}~\snm{Santi}\ead[label=e1]{lapo.santi@ucdconnect.ie}\orcid{0009-0005-9363-3353}},
\author[B]{\fnms{Nial}~\snm{Friel}\ead[label=e2]{nial.friel@ucd.ie}\orcid{0000-0003-4778-0254}}

\address[A]{School of Mathematics and Statistics,
University College Dublin, Dublin, Ireland\printead[presep={,\ }]{e1}}

\address[B]{Insight Centre for Data Analytics,
University College Dublin, Dublin, Ireland\printead[presep={,\ }]{e2}}
\end{aug}

\begin{abstract}
The Bradley--Terry model is widely used for the analysis of pairwise comparison data and, in essence, produces a ranking of the items under comparison. We embed the Bradley--Terry model within a stochastic block model, allowing items to cluster. The resulting Bradley--Terry SBM (BT--SBM) ranks clusters so that items within a cluster share the same tied rank. We develop a fully Bayesian specification in which all quantities---the number of blocks, their strengths, and item assignments---are jointly learned via a fast Gibbs sampler derived through a Thurstonian data augmentation. Despite its efficiency, the sampler yields coherent and interpretable posterior summaries for all model components. Our motivating application analyzes men's tennis results from ATP tournaments over the seasons 2000--2022. We find that the top 100 players can be broadly partitioned into three or four tiers in most seasons. Moreover, the size of the strongest tier was small from the mid-2000s to 2018 and has increased since, providing evidence that men's tennis has become more competitive in recent years.
\end{abstract}


\end{frontmatter}

\section{Introduction}

Rankings are ubiquitous. From institutional league tables in education and health to online reviews, sports tournaments and app ratings, ordered lists routinely inform decisions, allocate resources, and shape reputations. Their appeal lies in their simplicity: complex information about each item is distilled into a single number. Yet this simplicity can be misleading, creating a false sense of precision and granularity that the data often do not support. As \citet{goldstein1996} famously cautioned, there are \emph{“quantifiable uncertainties which place inherent limitations on the precision with which institutions can be compared”}–a critique that extends beyond schools and hospitals to rankings of all kinds.

Nowhere are these tensions more evident than in competitive sports, where rankings are both economically consequential and statistically fragile. In professional tennis -- the substantive focus of this paper–rankings are central. A seemingly minor drop in position can alter tournament seedings, jeopardize endorsement deals, affect a player's self-image, and ultimately determine whether a career remains sustainable \citep{Schottl2025tennis}. Rankings depend on players’ performances across tournaments, but it remains unclear to what extent a drop in rank reflects a genuine performance gap.

Ranking lists in many contexts, particularly sports, are typically strict, in the sense that if two players are selected at random, one player will always be ranked higher than the other. In so doing, such a ranking produces a strict complete ordering of the items (for example, players) under consideration and implies that tied ranks are not permitted. 

In professional tennis, this assumption can be particularly brittle. Two players separated by a single rank may never have competed head-to-head, may have followed markedly different tournament schedules, or may have been subject to different quality of opponents. 

A strict complete ordering nonetheless treats these noisy and context-dependent differences as meaningful, potentially overstating separation where none is credibly supported by the data.

The limitations of strict rankings have only recently begun to be addressed in Bayesian approaches \citep{PEARCE_2025, piancastelliClusteredMallowsModel2024}, which go beyond strict total orders by accommodating tied ranks of items. 

Building on this line of work, our objective is to develop a model that clusters items that share the same rank while imposing a strict ordering only on the cluster labels. 
This yields a coarser ranking, but it is often a more faithful representation of uncertainty: rather than forcing spurious pairwise distinctions, players of similar abilities may be assigned to the same cluster.

To operationalize this idea, we take the Bradley–Terry (BT) model \citep{bradleyRankAnalysisIncomplete1952} as our starting point. The BT framework assigns each player a latent strength (or ability) parameter. 

Sorting these parameters yields a ranking of players, thereby inducing a strict total order. In other words, the BT model inherits the very limitation we aim to overcome: it treats even negligible differences in estimated strengths as meaningful, producing a strictly ordered list of players.

To relax this assumption, we extend the BT framework by embedding it with the Stochastic Block Model (SBM) \citep{holland_stochastic_1983, Nowicki_2001}. This combination preserves the well-known BT likelihood while introducing a latent block structure that clusters players into groups of similar strength. The resulting model replaces spurious distinctions with a clustered ordering in which only the clusters themselves are strictly ranked.

This formulation, which combines pairwise comparison modelling with latent clustering, directly motivates the following methodological contributions: (i) \emph{End-to-end probabilistic analysis of pairwise-data networks:} we model pairwise outcomes as a directed comparison network and make \emph{ranking itself a random object}. The BT likelihood is retained, but ranking is transferred from individual items to \emph{ordered blocks}. The result is a fully generative Bayesian model in which posterior samples propagate uncertainty through every layer–from block membership to block strengths and even the number of blocks–so that summaries and decisions are \emph{probabilistic by design} rather than point-estimates; (ii) \emph{Finite yet data-driven number of blocks:} a nonparametric Gibbs-type (Gnedin) prior partitions items into a random but finite number of blocks, with reinforcement dynamics well suited to bounded, tournament-like populations such as seasonal sports leagues. 

This prior combines parsimony with heavy-tailed flexibility and lets $K$, the number of clusters, be learned from the data; (iii) \emph{Fully conjugate inference with variable dimension:} a Thurstonian data augmentation \citep{caronEfficientBayesianInference2012} yields closed-form full conditionals and a fast single-site Gibbs sampler that \emph{jointly} learns the number of blocks, their strengths, and player memberships–traversing a variable-dimensional posterior \emph{without} reversible-jump or split–merge moves, with per-iteration cost $\mathcal{O}(|E| + nK)$, where $E$ is the total number of observed interactions; and (iv) \emph{Uncertainty-aware posterior summaries:} from the same posterior draws, we obtain fully probabilistic and internally coherent summaries that directly address applied questions–the probability that a player belongs to a given tier, the distribution over $K$, odds-interpretable block strengths, and entropy-based measures of competitive balance with credible intervals. These posterior quantities are designed for reporting, comparison, and principled decision-making.

Empirically, we apply the BT–SBM to twenty-three seasons (2000/21 to 2022/23) of the results of matches from the \emph{Association of Tennis Professionals} (ATP), focusing on the top $n = 105$ players each year. The posterior summaries reveal a clear and mutually reinforcing narrative: during the so-called big four era (when the players, Federer, Nadal, Djokovic, and Murray dominated), from roughly 2008–2018, the elite field contracts and stabilises, producing a sharp concentration of strength, reflected in an small number of players estimated to be assigned to the strongest cluster of players. While after 2018, the distribution of posterior memberships broadens, signalling the rise of a more open competitive landscape. Beyond interpretability, the BT–SBM also improves predictive accuracy relative to the standard Bradley–Terry model, 

confirming the practical value of block-ranking as a robust, data-supported alternative to noisy individual rankings.

Taken together, our results make the case for moving beyond strict complete orderings toward clustered, uncertainty-aware rankings that offer a more interpretable and empirically faithful view of pairwise data.

\section{Mens APT results data}\label{sect:data}

The substantive application which motivates our work concerns mens professional tennis. Specifically, we consider results of matches from the \emph{Association of Tennis Professionals} (ATP) men's tour\footnote{See \url{https://www.atptour.com/} for further details.}. The APT men's tour comprises a series of tournaments. A feature of the APT series is that the performance of each player in tournament is then used to determine an APT rank of each player. As highlighted above, this is a strict complete ranking. We focus on an analysis of matches over $23$ seasons from $2000/2001$ through to $2022/2023$, where we use the 2017/2018 season as an illustrative example. Throughout this study period we fixed the number of players each season to $n=105$, to ensure comparability across seasons. 

For each season, we record the outcome of each match as an ordered pair between two players. In particular, let $w_{ij}$ denote the number of times that player $i$ defeated player $j$, for all $(i, j) \in E$, where $E = \{(i, j) : n_{ij} > 0\}$. While $n_{ij}$ denotes the total number of times players $i$ and $j$ played each other that season. 

We collect these results into a matrix $\mathbf{W}$ of size $n \times n$. Note that the matrix is not symmetric: if two players met $n_{ij}$ times, then $w_{ji} = n_{ij} - w_{ij}$. We define the match count matrix 
\(
\mathbf{N} = \mathbf{W} + \mathbf{W}^\top,
\)
so that its $(i,j)$th entry is given by
\(
n_{ij} = w_{ij} + w_{ji}, \; i,j = 1,\dots,n.
\)

By construction, $\mathbf{N}$ is symmetric and $n_{ii} = 0$ for all $i$.

\begin{figure}[htpb]
  \centering
  \includegraphics[width=\linewidth]{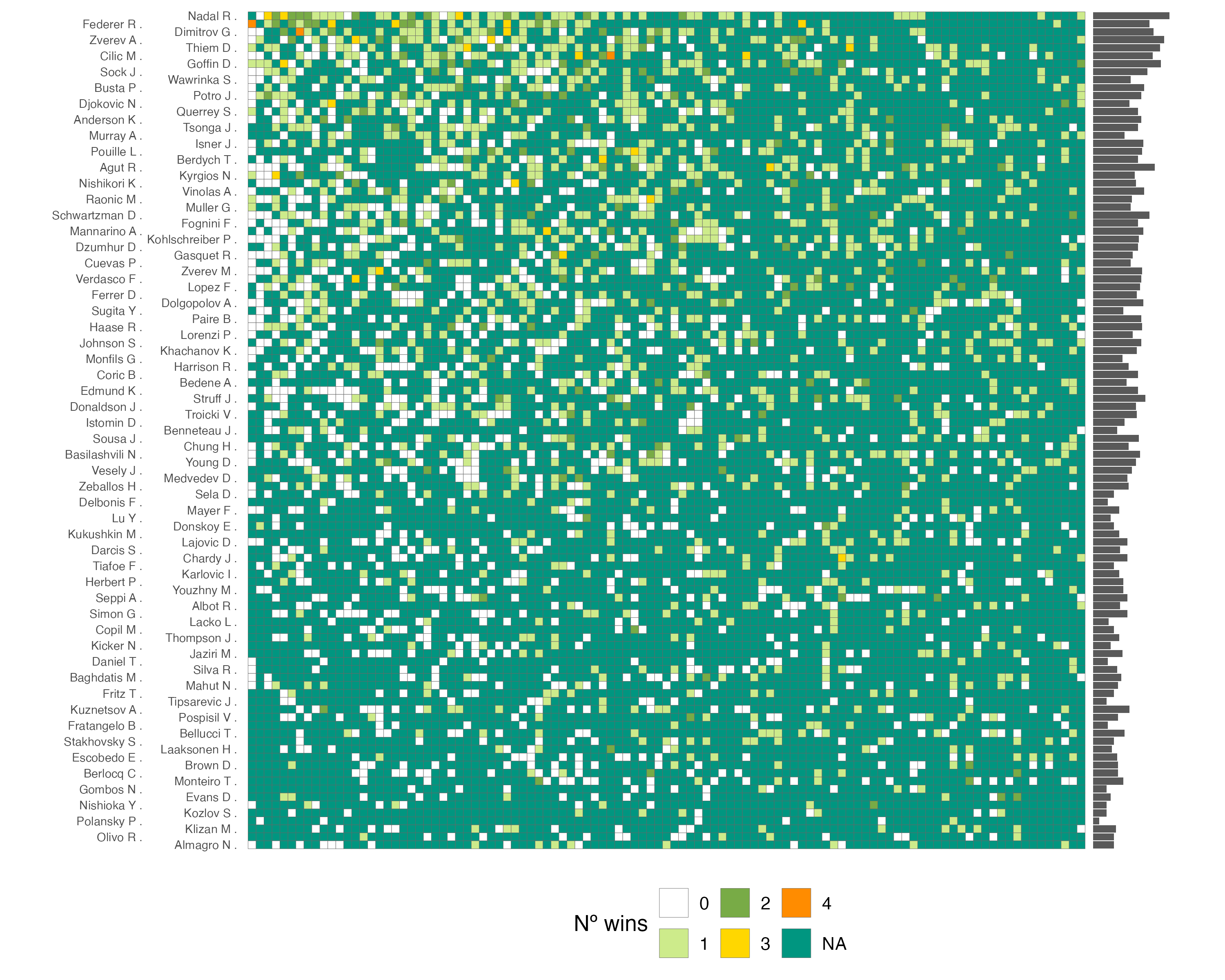}
  \caption{The match outcome matrix $\mathbf{W}$ for the ATP 2017-2018 season represented as an adjacency matrix where the players are ordered by their end-of-season ATP ranking. Each cell represents the number of wins of the row player against the column player. Dark green denotes missing data; white indicates no wins; yellow to orange indicates increasing win counts. Right margin bars show total matches played. The names on the y-axis are staggered across two columns to improve readability.}
  \label{fig:exploratory_plots}
\end{figure}

Figure~\ref{fig:exploratory_plots} presents a graphical representation of $\mathbf{W}$, with players ordered along both axes according to their ATP ranking at the end of the season $2018$. Each cell encodes the number of wins of the row player against the column opponent, using a colour gradient. Dark green indicates missing matches, white means no wins, and the light-green to orange palette represents increasing win counts. On the right, we show the total number of matches played by each player. 

Several structural features emerge from an inspection of Figure~\ref{fig:exploratory_plots}. The matrix is sparse -- most pairs of players never meet -- and this sparsity reflects the well-known “winners-play-more” effect of knockout tournaments: lower-ranked players (in the bottom-right corner) face each other less often, while top players (top-left) meet frequently.
The matrix is also skew-symmetric along the main diagonal. In the top-right corner, matches predominantly result in wins for higher-ranked players; in the bottom-left, we mainly observe losses. Along the diagonal band, by contrast, outcomes tend to be more balanced, with reciprocal results and local cycles that are inconsistent with the notion of a strict global ordering.

Just as important is what the pairwise comparison matrix fails to capture. Each cell simply counts wins but ignores contextual factors: the timing of the match (early vs. late season), the surface (clay, grass, hard), the physical condition of the player (injuries, fatigue), or situational aspects (home advantage, retirements). 
These structural features -- sparsity, skew-symmetry, and local inconsistencies -- motivate moving beyond a strict one-dimensional ordering to a model that explicitly accommodates uncertainty, allowing for tied or clustered ranks when the data do not credibly support fine distinctions.

\section{Literature Review}

We briefly review key contributions that inform our work. At the core of our framework lies the Bradley-Terry (BT) model, which assigns the probability that item \(i\) defeats item \(j\) as
\begin{equation}\label{eq:BTprob}
p(i \; \text{beats} \; j \mid \bm \lambda) = \frac{\lambda_i}{\lambda_i + \lambda_j}, 
\qquad \lambda_i, \lambda_j > 0 \quad \text{for all }(i,j) \in E,
\end{equation}
where \(\bm{\lambda} = \{\lambda_1, \ldots, \lambda_n\}\) denotes the vector of item-specific strength or ability parameters for the \(n\) items under comparison. The BT model remains a cornerstone of paired-comparison modelling and has inspired a rich body of work \citep{caronEfficientBayesianInference2012, glickmanBayesianLocallyOptimal2008, seymourBayesianSpatialBradleyTerry2021, wainerBayesianBradleyTerryModel2023, whelanPriorDistributionsBradleyTerry2017, croonLatentStructureModels1993}.

Ordering the parameters of \(\bm \lambda\) yields an implicit ranking: if \(\lambda_i > \lambda_j\), then the probability that \(i\) defeats \(j\) exceeds 0.5, and sorting the strengths produces a complete linear order of the items. The classical BT model therefore enforces a strict ranking through the latent strengths. Such order is characterized by the property of linear stochastic transitivity (LST): if \(\lambda_i > \lambda_j\) and \(\lambda_j > \lambda_k\), then necessarily \(\lambda_i > \lambda_k\), so all items can be placed on a single latent strength scale.

While convenient, LST is implausible in practice, as it assumes that outcomes of a match depend solely on the strength parameters of both player. In tennis, however, performances are shaped by multiple interacting factors, as discussed in Section~\ref{sect:data}. These sources of variability generate ranking-inconsistent outcomes, casting doubt on the adequacy of a representation of strength on a per-player basis.

To capture uncertainty and heterogeneity in the data, we move beyond strict one-to-one rankings by allowing players to be grouped into clusters. In our setting, players in the same cluster share the same strength parameter \(\bm \lambda\). This implies that if two players belong to the same cluster, the probability of one defeating the other is exactly 0.5. Clustering thus provides a principled way to model rank-indifference: groups of players are indistinguishable in terms of estimated strength, while different clusters can still be ordered according to their common \(\bm \lambda\).

Several authors have extended BT models to incorporate clustering. \citet{Wu_2015} propose a mixture BT model to capture population heterogeneity, grouping items with distinct preference patterns, though without explicitly modeling rank indifference. \citet{hsiaoJointClusteringRanking2021} consider a “Cluster-BT” framework in which items are partitioned into latent clusters and within-cluster comparisons follow the BT model, while inter-cluster comparisons are treated as random/noise (i.e. non-informative).
\citet{Spearing_2023} introduced a BT-based model that clusters both items and interactions (e.g., games or matchups) to account for inconsistencies in outcomes, in the context of basketball data. In contrast, our focus is strictly on item-level clustering: we aim to discover interpretable transitive structures by estimating a partition of tied-ranked players.
\citet{PEARCE_2025} developed a rank-clustering approach in the BT family that fuses nearby strengths using a spike-and-slab prior. While their model shares our motivation of relaxing strict complete orderings, it treats clustering as a form of shrinkage on strength parameters. 

In contrast to these papers, we adopt a fully generative Bayesian perspective that partitions items into latent clusters of comparable strength. Within each cluster, players are treated as rank-indifferent, while clusters themselves are strictly ordered. This structure balances interpretability with flexibility: it acknowledges the limits of LST at the player level while retaining a coherent transitive hierarchy at the cluster level. Recovering this clustered ordering and quantifying its uncertainty is the central goal of our model.

\section{Embedding the BT model within an SBM}
\label{sec:aug_BT_SBM}

Competitions, whether in football, chess, or tennis, are inherently relational systems where hidden hierarchies and clusters emerge from repeated encounters. SBMs have been successfully employed to uncover such latent structures in various sports. Notably, \citet{basiniAssessingCompetitiveBalance2023} apply SBMs to football leagues to assess competitive balance, while \citet{SystematicVacaRam2022} review applications to chess tournaments and American football. In a similar spirit, we employ SBMs to model tennis competitions. Building on this line of work, we introduce the Bradley–Terry Stochastic Block Model (BT–SBM), which combines the Bradley–Terry framework for pairwise comparisons with the SBM for clustering relational data. Before detailing the model, we briefly review the main features of SBMs.

\subsection{Overview of SBMs}
The SBM provides a flexible probabilistic framework for latent clustering in networks. Each node \( i \in \mathcal \{1,\ldots,n \) is assumed to belong to an unobserved block \( x_i \in \{1, \dots, K\} \), and the presence of an edge between nodes depends solely on their block memberships – a property known as \emph{stochastic equivalence}. We first describe the canonical SBM \citep{Nowicki_2001}, before reviewing some recent approaches in this area. We begin by letting \( \mathbf{x} = (x_1, \ldots, x_n) \) denote the latent block assignments, $y_{ij}$, the binary outcome of whether an edge is present or not is modelled as,  
\[
y_{ij} \mid \mathbf{x}, \bm{\theta} \sim \mathrm{Bernoulli}(\theta_{x_i x_j}),
\]
where \( \bm{\theta} = (\theta_{rs})_{r,s=1}^K \) is a \( K \times K \) block–connectivity matrix. Where \(\theta_{rs}\) represents the probability of an edge between any node in block \(r\) and any node in block \(s\). Each element \(\theta_{rs}\) is typically assigned an independent \(\mathrm{Beta}(a,b)\) prior.  
The assignment vector \(\mathbf{x}\) is modelled via block mixing proportions \(\bm{\pi} = (\pi_1, \dots, \pi_K)\), drawn from a Dirichlet prior, \(\bm{\pi} \sim \mathrm{Dirichlet}(\alpha_1, \dots, \alpha_K)\). Marginalizing over \(\bm{\pi}\) induces a Dirichlet–multinomial distribution on \(\mathbf{x}\), favouring balanced cluster allocations when the hyperparameters \(\alpha_k\) are symmetric.

Recent developments in Bayesian nonparametrics have introduced the overarching class of partition priors for $\mathbf{x}$, known as Gibbs–type priors, which encompass the traditional Dirichlet–multinomial, as outlined above, as a special case \citep{Pitman1996,DeBlasi2015}. Examples include the Gnedin process \citep{Gnedin2006}, along with the well-known Dirichlet process \citep{Ferguson1973} and the Pitman–Yor process \citep{PitmanYor1997}. A key virtue of Gibbs–type priors is that they allow the number of occupied clusters \(K\) to be data-driven while offering fine control over the dynamics of reinforcement (how cluster sizes grow) and innovation (how new clusters are created) \citep{DeBlasi2015,deblasi2013}.

\subsection{A Gnedin Prior for the Latent Partition}\label{sect:priorGN}

We now replace the simpler Dirichlet–multinomial prior on \(\mathbf{x}\) by a Gibbs–type prior, thereby linking smoothly with the earlier mention of mixing weights and allowing more flexibility in the number of clusters. Concretely, we posit  
\begin{equation}\label{eq:gibbs_prior}
p(\mathbf{x}) = \psi_{n, K(\mathbf{x})} \,\prod_{k=1}^{K(\mathbf{x})} (1 - \sigma)_{\,m_k - 1}, 
\quad \sigma < 1,
\end{equation}
where \(m_k = |\{i: x_i = k\}|\) is the size of block \(k\), \(K(\mathbf{x})\) is the number of unique labels in the vector $\mathbf x$, and \((a)_r\) denotes the ascending factorial (Pochhammer) \((a)_r = a (a+1)\cdots(a + r - 1)\). From now on, we will simply refer to $K(\mathbf x)$ as $K$ for ease the notation, while aware that $\mathbf x$ encodes the number of clusters $K$. 

The triangular recursion on \(\psi\),
\[
\psi_{n,K} = (n - \sigma K)\,\psi_{n+1,K} + \psi_{n+1,K+1}, 
\quad \psi_{1,1} = 1,
\]
induces the predictive urn scheme \citep{MR2370077}:
\[
p(x_{n+1} = k \mid \mathbf{x}) \propto
\begin{cases}
\psi_{n+1,K}\,(m_k - \sigma), & k = 1, \dots, K,\\
\psi_{n+1,K+1}, & k = K+1.
\end{cases}
\]

Within this class, we choose the \emph{Gnedin model} \(\mathbf{x} \sim p_{\mathrm{GN}}(\mathbf x \mid \gamma)\), corresponding to \(\sigma = -1\) and a single parameter \(\gamma \in (0,1)\). In that case,
\[
\psi_{n,K} = \frac{(\gamma)_{\,n-K}\,\prod_{k=1}^{K-1}(k^2 - \gamma k)}{\prod_{i=1}^{n-1}(i^2 + \gamma i)},
\]
and the predictive rule simplifies to:
\begin{equation}\label{eq:gnedin_prior}
    p(x_{n+1} = k \mid \mathbf{x}) \propto (m_k + 1)\,(n - K + \gamma), \qquad
p(x_{n+1} = K+1 \mid \mathbf{x}) \propto K^2 - K\gamma.
\end{equation}

In the context of our motivating application to tennis data, we favour the Gnedin prior, for several reasons which we now outline. In modelling latent player groups, the true number of blocks is unknown but plausibly finite and moderate (not diverging with $n$), because only a finite set of players competes in a season. Relative to the Dirichlet–multinomial discussed above, the Gnedin prior allows a \emph{random yet almost surely finite} number of clusters \(K\), thereby preserving flexibility \citep[see, e.g.,][]{DeBlasi2015,Legramanti_2022}. 

In particular, \citet{Legramanti_2022} emphasise that the Gnedin process generates random and finite number of groups. But also that it is subject to what they term a reinforcement process.

This aligns well with our tennis context, where ``reinforcement'' captures the idea that the prior favours the expansion of existing clusters, giving additional weight to groups that are already established, rather than continuously spawning new ones. Finally, in the same paper they observe that the prior on the \emph{population} number of groups has mode at \(1\) and heavy tails, favouring parsimonious yet robust specifications of \(K\).

Another useful property is that one can study the prior distribution of \(K\) in \emph{closed analytic form}. 
Specialising the general Gibbs–type result \eqref{eq:gibbs_prior} to the Gnedin case \((\sigma=-1)\) yields the exact pmf
\[
p(K = k \mid n,\gamma)
= \binom{n}{k}\,
\frac{(1-\gamma)_{k-1}\,(\gamma)_{\,n-k}}{(1+\gamma)_{\,n-1}},
\qquad k=1,2,\dots,n.
\]
Hence,
\begin{equation}\label{eq:expected_valueK}
\mathbb{E}[K \mid n,\gamma]
= \sum_{k=1}^{n} k\,p(K=k \mid n,\gamma)
= \frac{\Gamma(n+1)\,\Gamma(1+\gamma)}{\Gamma(n+\gamma)},
\end{equation}
which coincides with the expression reported in \citet{Legramanti_2022}. 
In addition, the variance of \(K\) admits a closed form:
\begin{equation}\label{eq:varianceK}
\operatorname{Var}(K \mid n,\gamma)
= \mathbb{E}[K \mid n,\gamma]\,\bigl[n-\gamma(n-1)\bigr]
- \mathbb{E}[K \mid n,\gamma]^2.
\end{equation}

This compact expression shows how the variability of the number of clusters depends jointly on \(n\) and the hyperparameter \(\gamma\). 
Lower values of \(\gamma\) inflate the term \(n - \gamma(n-1)\), producing a heavier-tailed prior on \(K\) and thereby increasing the chance of allocating additional clusters even under moderate sample sizes. A more detailed discussion of the mean and the variance of $K$ is reported in Appendix~\ref{sect:mean_var_K_gnedin}.

From an empirical perspective, recent studies within the SBM/eSBM literature report strong performance of Gnedin–based specifications relative to other Gibbs–type competitors \citep{Legramanti_2022,Lu_2025}.

\subsection{Block-clustered BT likelihood}

Once the latent partition $\mathbf{x} \sim p_{\mathrm{GN}}(\mathbf x \mid \gamma)$ is drawn and the number of occupied blocks $K$ determined, each block is associated with a single latent \emph{strength} parameter $\lambda_k$. Intuitively, $\lambda_k$ represents the overall competitive strength of players in block $k$, capturing how likely they are to prevail against members of other groups. The collection of block strengths thus forms a $K$-dimensional vector $\bm{\lambda} = (\lambda_1, \ldots, \lambda_K)$, which summarises the structure of the competition at the group level. To retain flexibility while maintaining conjugacy with the likelihood, we assign each $\lambda_k$ an independent Gamma prior,
\begin{equation}\label{eq:strength_prior}
\lambda_k \mid \mathbf{x} \sim \Gamma(a,b),
\end{equation}
a standard conjugate choice in Bradley–Terry or Plackett–Luce models \citep{gormleyGradeMembershipModel2009,guiverBayesianInferencePlackettLuce2009}. This choice accommodates a wide range of relative block strengths through its shape and rate parameters, while preserving interpretability: higher values of $\lambda_k$ correspond to stronger groups. Conditional on $\mathbf{x}$ and $\bm{\lambda}$, the probability that player \(i\) beats \(j\) is
\[
p(i \text{ beats } j \mid \mathbf{x},\bm{\lambda}) = \frac{\lambda_{x_i}}{\lambda_{x_i} + \lambda_{x_j}},
\]
and hence the full likelihood over all comparisons is
\begin{equation}
\mathcal{L}(\mathbf{W} \mid \bm{\lambda}, \mathbf{x}, \mathbf{N})
= \prod_{i<j: n_{ij}>0}
\left(\frac{\lambda_{x_i}}{\lambda_{x_i} + \lambda_{x_j}}\right)^{w_{ij}}
\left(\frac{\lambda_{x_j}}{\lambda_{x_i} + \lambda_{x_j}}\right)^{n_{ij} - w_{ij}}.
\label{eqn:BT-likelihood}
\end{equation}
\subsection{Data augmentation and conjugacy}
\label{sec:augmentation}

Based on the development of the BT-SBM as outlined so far, one could focus on the posterior distribution,
\begin{align*}
p\Bigl(\bm{\lambda},\mathbf{x}, \mid \mathbf{W},  \mathbf{N} \Bigr) \propto &\;
\underbrace{\mathcal{L}(\mathbf{W}\mid \bm{\lambda},\mathbf{x},  \mathbf{N})}_{\text{BT- likelihood as in \eqref{eqn:BT-likelihood}}}
\times \underbrace{p(\bm{\lambda} \mid  \mathbf{x})}_{\text{Strengths' prior as in \eqref{eq:strength_prior}}}
\times \underbrace{p_{\mathrm{GN}}(\mathbf x \mid \gamma)}_{\text{Gnedin prior spec. of \eqref{eq:gibbs_prior}}}.
\end{align*}

However, posterior inference, via a standard Metropolis-within-Gibbs is not straightforward. In particular, this results in a slow mixing chain that converges slowly to the posterior distribution. 

To address this issue, we follow the approach presented in \citet{caronEfficientBayesianInference2012} that reinterprets each individual comparison as a continuous-time event, by introducing the latent variables:
\[
Y_{k,i} \sim \mathrm{Exp}(\lambda_{x_i}), \quad Y_{k,j} \sim \mathrm{Exp}(\lambda_{x_j}),
\quad k = 1, \dots, n_{ij}.
\]
In this formulation, often referred to as \emph{Thurstonian}, player \(i\) “wins” the \(k\)th match when \(Y_{k,i} < Y_{k,j}\), which yields
\(
p(Y_{k,i} < Y_{k,j}) = \dfrac{\lambda_{x_i}}{\lambda_{x_i} + \lambda_{x_j}},
\)
coinciding with the usual BT probability that player $i$ beats player $j$. We then define
\[
Z_{ij} = \sum_{k=1}^{n_{ij}} \min\{Y_{k,i}, Y_{k,j}\} \;\sim\; \Gamma\bigl(n_{ij},\, \lambda_{x_i} + \lambda_{x_j}\bigr),
\]

This results in a complete-data likelihood which is expressed as
\begin{equation}\label{eq:single_product_likelihood}
\mathcal{L}(\mathbf{W}, \mathbf{Z} \mid \bm{\lambda}, \mathbf{x}, \mathbf{N})
= \prod_{i<j: n_{ij}>0}
\lambda_{x_i}^{\,w_{ij}} \,
\lambda_{x_j}^{\,n_{ij}-w_{ij}}
\,\frac{Z_{ij}^{\,n_{ij}-1}}{\Gamma(n_{ij})}
\exp\!\bigl[-(\lambda_{x_i} + \lambda_{x_j})Z_{ij}\bigr].
\end{equation}
The result of augmenting the likelihood with $\mathbf{Z}$ is that this allows one to develop a Gibbs sampler for all parameters in the model, as we will shortly outline. Moreover, the resulting MCMC algorithm is fast to run and offer the potential to scale well to larger datasets. 

\subsection{Full Posterior Distribution}
\label{sec:full_posterior}

Combining the above elements, we obtain the full posterior, 

\begin{align*}
p\Bigl(\bm{\lambda},\mathbf{x},\mathbf{Z} \mid \mathbf{W},  \mathbf{N} \Bigr) \propto &\;
\underbrace{\mathcal{L}(\mathbf{W},\mathbf{Z}\mid \bm{\lambda},\mathbf{x},  \mathbf{N})}_{\text{Augmented likelihood as in \eqref{eq:single_product_likelihood}}}
\times \underbrace{p(\bm{\lambda} \mid  \mathbf{x})}_{\text{Strengths' prior as in \eqref{eq:strength_prior}}}
\times \underbrace{p_{\mathrm{GN}}(\mathbf x \mid \gamma)}_{\text{Gnedin prior spec. of \eqref{eq:gibbs_prior}}}.
\end{align*}

Substituting the full expressions for each term on the right hand side, and letting the proportionality sign absorb all the constants that do not depend on $ \{\bm \lambda, \mathbf x, \mathbf Z\}$ yields,

\begin{align}
p\Bigl(\bm{\lambda},\mathbf{x},\mathbf{Z}\mid \mathbf{W},  \mathbf{N}\Bigr) \propto \; & \prod_{1\le i<j\le n: \, n_{ij}>0} \frac{\lambda_{x_i}^{\,w_{ij}}\,\lambda_{x_j}^{\,n_{ij}-w_{ij}}}{\Gamma(n_{ij})}\,Z_{ij}^{\,n_{ij}-1}\,\exp\Bigl\{-(\lambda_{x_i}+\lambda_{x_j})Z_{ij}\Bigr\} \nonumber\\[2mm]
& \times \prod_{k=1}^{K} \frac{b^a}{\Gamma(a)} \lambda_k^{\,a-1}\exp(-b\lambda_k)
\times \psi_{n,K}\prod_{k=1}^{K}(1-\sigma)_{\,m_k-1}.
\label{eq:full_posterior_detailed}
\end{align}

This compact form is the basis for deriving the full conditional updates in Gibbs sampling or other posterior explorations.

In summary, Figure~\ref{fig:bt_sbm_dag} provides a joint representation of the augmented BT--SBM model and its generative process.

\begin{figure}[htpb]
\centering
% ---- Left: DAG ----
\begin{minipage}[t]{0.48\linewidth}
\centering
\begin{adjustbox}{max width=\linewidth, valign=t}
\begin{tikzpicture}[node distance=1.6cm, auto,>=stealth,
  obs/.style={draw, rectangle, rounded corners=2pt},
  lat/.style={draw, circle},
  det/.style={draw, circle, dashed}
]
 % Hyperparameters
 \node[lat] (a)  {\(a\)};
 \node[lat, right=1cm of a] (b)  {\(b\)};
 \node[lat, left=6.5cm of a] (sigma) {\(\sigma\)};
 
 % x_i, x_j
 \node[lat, below=1.6cm of sigma]    (x_i) {\(x_i\)};
 \node[lat, right=0.8cm of x_i]      (x_j) {\(x_j\)};

 % K = K(x)
 \node[det, right=1.2cm of x_j]      (K) {\(K=K(\mathbf x)\)};

 % lambda plate (k=1..K)
 \node[lat, right=1.2cm of K]        (lambda) {\(\lambda_k\)};
 \node[draw, rectangle, fit=(lambda), inner sep=0.45cm, 
       label=below right:{\(k=1,\dots,K\)}] (plateK) {};

 % Plate for items
 \node[draw, rectangle, fit=(x_i)(x_j), inner sep=0.45cm, 
       label=below left:{\( (i,j) \in E\)}] (plateN) {};

 % Observed N_ij and W_ij and latent Z_ij
 \node[obs, below left=2.8cm and 1cm of K, fill=gray!50] (W) {\(w_{ij}\)};
 \node[obs, below=1cm of W, fill=gray!50]   (N) {\(n_{ij}\)};
 \node[lat, right=2.0cm of W]        (Z) {\(Z_{ij}\)};
 % Plate for pairs i<j
 \node[draw, rectangle, fit=(N)(W)(Z), inner sep=0.35cm, 
       label=below:{\(i<j\)}]  (platePairs) {};

 % Arrows
 \draw[->] (a)   -- (lambda);
 \draw[->] (b)   -- (lambda);
 \draw[->] (sigma) -- (x_i);
 \draw[->] (sigma) -- (x_j);
 \draw[->] (x_i)  -- (W);
 \draw[->] (x_j)  -- (W);
 \draw[->] (lambda) -- (W);
 \draw[->] (x_i)  -- (Z);
 \draw[->] (x_j)  -- (Z);
 \draw[->] (lambda) -- (Z);
 \draw[->] (N) -- (W);
 \draw[->] (N) -- (Z);
 \draw[->] (plateN) -- (K);
 \draw[->] (K)   -- (plateK);
\end{tikzpicture}
\end{adjustbox}
\label{fig:bt_sbm_dag2}
\end{minipage}\hfill
% ---- Right: Algorithm (no float inside) ----
\begin{minipage}[t]{0.51\linewidth}
\vspace{0pt}
\captionsetup{type=algorithm}
\begin{algorithmic}[1]
\REQUIRE \(n\), \(\mathbf N=\{n_{ij}\}_{i<j}\), \(a,b,\gamma\).
\STATE \textbf{Initialize:} $\{K, x_1\} \gets 1, \lambda_1\sim\Gamma(a,b)$;
\FOR{$i=2$ \TO $n$}
  \STATE $x_i \sim \mathrm{GN}(\gamma)$ as in \eqref{eq:gnedin_prior}
  \IF{$x_i=K+1$}
    \STATE $K\gets K+1$; $\lambda_{K}\sim\Gamma(a,b)$
  \ENDIF
\ENDFOR
\FORALL{pairs \(i<j\) with \(n_{ij}>0\)}
  \STATE $w_{ij} \mid n_{ij}, \mathbf x, \bm \lambda \sim \mathrm{Bin}\!\left(n_{ij},\, \frac{\lambda_{x_i}}{\lambda_{x_i}+\lambda_{x_j}}\right)$
\ENDFOR
\end{algorithmic}
\label{alg:generating}
\end{minipage}
\caption{On the left, the directed acyclic graph (DAG) highlights the hierarchical structure of the model: 
hyperparameters \((a,b)\) and \(\sigma)\) govern the latent cluster strengths \(\bm\lambda\) and the item-specific allocations \(\mathbf x\), respectively. 
In turn, these latter determine the distribution of the observed match outcomes \(\mathbf W = (w_{ij})\) given the number of matches \(\mathbf N = (n_{ij})\). The number of occupied blocks \(K\) is determined by \(\mathbf x\), and \(K\) affect the size of the plate for \(\lambda_k\). The grey-coloured boxes represent observed outcomes.
On the right, the algorithm sketches the generative procedure required to sample from the BT-SBM. 
Items are allocated one at a time either to an existing block or to a newly created one according to the predictive distribution 
\(p(x_{n+1}\mid \mathbf x)\) in Eq.~\ref{eq:gnedin_prior}, while the number of occupied blocks \(K\) evolves accordingly. 
After block assignments and cluster strengths have been sampled, match outcomes are generated conditionally on $(\mathbf{x},\bm \lambda, \mathbf N)$. This procedure yields partitions with the characteristic features of the Gnedin prior–namely, a small modal number of clusters and a heavy-tailed distribution–and match outcomes reflecting $K$ different tiers of strength.
\label{fig:bt_sbm_dag}}
\end{figure}

\section{Posterior inference and single-site Gibbs sampling}
\label{sec:posterior_inference}

Having derived the full posterior distribution in Section~\ref{sec:full_posterior}, we now turn to inference. In particular, we employ a single-site Gibbs sampler which exploits the conditional conjugacy of the augmented model, which leads to closed-form full-conditional distributions for all variables in (\ref{eq:full_posterior_detailed}), which we now detail.

%------------------------------------------------------------------------
\paragraph*{Auxiliary--variable update:}
\label{subsec:Z_update}

For every pair with at least one match ($n_{ij}>0$),
\begin{equation}
 Z_{ij}\mid\text{rest}\;\sim\;
 \Gamma\!\left(n_{ij},\,
    \lambda_{x_i}+\lambda_{x_j}\right).
 \label{eq:z_update}
\end{equation}

%------------------------------------------------------------------------
\paragraph*{Block--strength update:}
\label{subsec:lambda_update}

Let $I_k=\{i:x_i=k\}$ and set
\[
 w_i=\sum_{j\ne i}w_{ij},
 \qquad 
 Z_i=\sum_{j\ne i}Z_{ij}.
\]
With a $\Gamma(a,b)$ prior, the full conditional is
\begin{equation}
 \lambda_k\mid\text{rest}
 \;\sim\;
 \Gamma\bigl(
   a+\!{\textstyle\sum_{i\in I_k}} w_i,\;
   b+\!{\textstyle\sum_{i\in I_k}} Z_i
   \bigr).
 \label{eq:lambda_update}
\end{equation}
See Appendix~\ref{sect:appendix_derivations} for further details of this derivation.

%--------------------------------------------------------------------
\paragraph*{Block assignment update:}
\label{subsec:x_update}

For each item $i$ we sample the label $x_i$ from its unnormalised full
conditional
\begin{equation}
 p(x_i = k \mid \text{rest})
 \;\propto\;
 p(x_i = k \mid \mathbf x_{-i})
 \;\times\;
 \frac{p\bigl(\mathbf W \mid x_i = k,\, \mathbf x_{-i},
        \bm\lambda,\,\mathbf Z\bigr)}
    {p\bigl(\mathbf W_{-i} \mid \mathbf x_{-i},
        \bm\lambda,\,\mathbf Z\bigr)},
 \label{eq:assignment_generic}
\end{equation}
where the prior term follows the urn scheme
shown in Sect.\ref{sect:priorGN}. In the second term, we have the ratio where the denominator cancels out those pairs that do not involve $i$. 

For convenience, we denote by $L_i(k)$ the likelihood factor associated with assigning item $i$ to cluster $k$, i.e.\ the term multiplying the prior in~\eqref{eq:assignment_generic}, defined as:
\[
 L_i(k)
 \; \coloneqq \;\frac{p\bigl(\mathbf W \mid x_i = k,\, \mathbf x_{-i},
        \bm\lambda,\,\mathbf Z\bigr)}
    {p\bigl(\mathbf W_{-i} \mid \mathbf x_{-i},
        \bm\lambda,\,\mathbf Z\bigr)} \;=\;
 \lambda_k^{\,w_i}\,
 e^{-\lambda_k Z_i}, \text{for } k = 1,\ldots K,
\]
with $w_i=\sum_{j\neq i}w_{ij}$ and
$Z_i=\sum_{j\neq i}Z_{ij}$.
(See App.~A\ref{sect:appendix_derivations} for details.)

For a new cluster $(k=K+1)$, we follow \citet{neal2000markov}'s Algorithm~3, by integrating out the yet--to--be-sampled $\lambda_{K+1}$ to obtain:
\begin{align}\label{eq:new_cluster_prob}
 L_i (k)
 &\;=\;
 \int_0^{\infty}
    \lambda_k^{w_i}\,e^{-\lambda_k Z_i}\,
    \frac{b^{a}}{\Gamma(a)}\,
    \lambda_k^{a-1}e^{-b\lambda_k}\,d\lambda_k \nonumber \\
 &\;=\;
 \frac{b^{a}\,\Gamma(a+w_i)}{\Gamma(a)}
 \,(b+Z_i)^{-(a+w_i)} \quad \text{for } k = K+1.
\end{align}

Combining prior and likelihood terms gives
\begin{equation}
 p(x_i = k \mid \text{rest})
 \;\propto\;
 \begin{cases}
  (m_k + 1)(n - K + \gamma)\;
  L_i(k),
  & k = 1,\dots,K;\\[10pt]
  (K^{2} - K\gamma)\;
  L_i (k),
  & k = K+1.
 \end{cases}
 \label{eq:assignment_fc}
\end{equation}
Normalising over $k\in\{1,\dots,K+1\}$ yields the sampling
probabilities. 
If the draw selects the new label $K\!+\!1$, we set $x_i \leftarrow K+1$, and then we sample
\[
 \lambda_{K+1} \sim \Gamma\bigl(a + w_i,\, b + Z_i\bigr),
\]
and increment the cluster count $K \leftarrow K + 1$.
Conversely, $K$ decreases naturally whenever a cluster becomes empty. Note that the number of \emph{occupied} clusters $K$ is a random quantity, determined by the current configuration of $\mathbf{x}$. 
Unlike reversible-jump or split-merge samplers, this approach explores a variable-dimensional posterior space without explicit trans-dimensional proposals: clusters are created or removed naturally as labels are updated.

\subsection{Hyperparameter specification and scale alignment}
\label{subsec:hyperparam_choice}

The behaviour of the model is chiefly governed by three hyperparameters: $(a,b)$, which define the prior over block strengths, and $\gamma$, which controls the partition structure and the overall model complexity (see Sect.~\ref{sect:priorGN}). Each block strength $\lambda_k$ is assigned a $\Gamma(a,b)$ prior (Eq.~\eqref{eq:strength_prior}), where the shape parameter $a$ and the rate $b$ affect the $\bm \lambda$ mean and variance, and also affect the probability of sampling a new cluster (see Eq.~\ref{eq:new_cluster_prob}).
Choosing appropriate values for $(a,b)$ is therefore essential, and we adopt a simple yet effective heuristic to fix them in a principled way. 

This heuristic is motivated by several considerations, the first of which arises from the multiplicative invariance of the Bradley--Terry likelihood, which depends only on ratios such as
\[
\frac{\lambda_{x_i}}{\lambda_{x_i}+\lambda_{x_j}}
=
\frac{c\,\lambda_{x_i}}{c\,\lambda_{x_i}+c\,\lambda_{x_j}}, \quad \text{for each} \, (i,j) \in E.
\]
\noindent
Therefore, the overall magnitude of $\bm{\lambda}$ cannot be identified from the data. To ensure identifiability, we normalise the strength vector $\bm{\lambda}$ at each iteration (step~5 in Alg.~\ref{alg:fullsampler}) through a standard procedure known as \emph{global rescaling}. Specifically, we impose that the arithmetic mean of the log-strengths be zero:
\[
\frac{1}{K}\sum_{k=1}^K \log \lambda_k = 0,
\]
as working on the log-scale provides greater numerical stability\footnote{This also allows one to interpret $\lambda_k$ as the odds of beating a block of average strength \citep{Newman2022}. For instance, a block with win probability $p_1$ against the average block has $\lambda_k = p_1 / (1-p_1)$, directly expressing strength in odds form.}. As a consequence, the rescaled $\log \bm{\lambda}$ satisfies, a posteriori, $\mathbb{E}[\log \lambda_k \mid \mathbf W] = 0$. Therefore, it seems a natural choice to select hyperparameters $(a,b)$ such that the prior mean of $\log \lambda_k$ aligns with the normalisation-induced zero expectation:
\[
\mathbb{E}_{\text{prior}}[\log \lambda_k \mid a,b]
=
\mathbb{E}[\log \lambda_k \mid \mathbf W]
=
0,
\]
which, for $\lambda_k \sim \Gamma(a,b)$, is obtained by setting
\[
b = \exp\{\psi(a)\},
\]
where $\psi$ is the derivative of the logarithm of the Gamma function, known as the \emph{digamma} function \citep{joramsoch2025}.

A second reason motivating this heuristic in choosing $(a,b)$ lies in its computational convenience, as it stabilises the propensity to sample new clusters. In particular, substituting $b = \exp\{\psi(a)\}$ in Eq.~\eqref{eq:new_cluster_prob} makes this probability less sensitive to variations in the scale of $\bm{\lambda}$. A detailed derivation and empirical validation of this argument are provided in Appendix~\ref{app:role_prior_scale_clean}.

Once $b = \exp\{\psi(a)\}$ has been fixed, the remaining shape parameter~$a$ is the only degree of freedom and governs the variance of $\log \lambda_k$, given by
\[
\mathrm{Var}(\log \lambda_k \mid a,b) = \psi_1(a),
\]
where $\psi_1$ is the \emph{trigamma} function, i.e., the derivative of the digamma function. From the relationship between $a$ and $\mathrm{Var}(\log \lambda_k)$ examined in Appendix~\ref{app:role_prior_scale_clean}, we find that values of $a \in [2,4]$ provide a reasonable balance, allowing for a moderate variability among the $\lambda_k$’s that can be quantified as:
\[
\tau = \sqrt{\psi_1(a)} = \mathrm{SD}(\log \boldsymbol{\lambda}) \in [0.53, 0.80].
\]

Having fixed $(a,b)$ to control the distribution of block strengths, we now turn to the parameter $\gamma$, which directly controls the mean and the variance of $K$ through the Gnedin prior \eqref{eq:gibbs_prior}. To adopt a conservative specification, we fix $\gamma = 0.8$, which for $n = 105$ yields
$\mathbb{E}[K \mid \gamma = 0.8, n = 105] \approx 2$ and 
$\operatorname{Var}(K \mid \gamma = 0.8, n = 105) \approx 46$, computed using \eqref{eq:expected_valueK} and \eqref{eq:varianceK}. 
This choice produces considerable shrinkage on $K$, as it discourages partitions with many clusters unless they are strongly supported by the data. However, the heavy right tail of the distribution still allows the sampler to explore different regions of the complex partition space.

\paragraph*{Code validation and computational complexity}

Before turning to real data, we validate both the statistical accuracy and the computational efficiency of the proposed algorithm. 
First, we evaluate its ability to recover the model parameters and overall structure through a simulation study (see App.~\ref{sec:simulation_appendix}). 
The results indicate that the posterior samples obtained via Alg.~\ref{alg:fullsampler} accurately recover both the true number of clusters ($K$) and the underlying partition structure ($\mathbf{x}$) when the data are generated from the BT--SBM, confirming the correctness of the inference procedure and the reliability of its implementation. 
Further details on the data-generating process, MCMC settings, and performance metrics are provided in Appendix~\ref{sec:simulation_appendix}.

In addition to statistical validation, we assess the computational scaling of the algorithm. 
Each Gibbs sweep requires \(\mathcal{O}(|E|)\) operations to update the latent variables \(Z_{ij}\), 
\(\mathcal{O}(nK)\) operations to update the cluster assignments, and \(\mathcal{O}(K)\) operations 
to update the block intensities \(\lambda_k\). 
Hence, the total cost per iteration scales as
\[
\mathcal{O}(|E| + K(n+1)),
\]
which is effectively linear in \(|E|\) in our setting, since the number of occupied blocks \(K\) 
is much smaller than the number of nodes (\(K \ll n\), typically \(K \leq 6\)). 

We verify this scaling empirically by simulating networks of increasing size \(n\) from a BT--SBM with \(K=5\) blocks (see App.~\ref{sec:simulation_appendix}).  
We control the expected sparsity of each network through its \emph{edge density}, defined as the ratio between the number of observed edges and the total number of possible directed edges, i.e. $\frac{|E|}{n(n-1)}$. Table~\ref{tab:complexity} reports the total runtime (in minutes) required for 10,000 Gibbs iterations on each of the simulated networks:

\begin{table}[htpb]
\centering
\rowcolors{1}{white}{gray!10}
\begin{tabular}{rrrr}
\toprule
$n$ & $|E|$ & Density & Total time (min) \\
\midrule
100   & 4,950   & 0.50 & 0.35 \\
500   & 24,950  & 0.10 & 2.06 \\
1,000 & 49,950  & 0.05 & 4.91 \\
5,000 & 249,950 & 0.01 & 123.07 \\
\bottomrule
\end{tabular}
\caption{Empirical scaling of total runtime with network size for 10,000 Gibbs iterations. 
Networks are generated from a BT--SBM with \(K=5\) blocks and varying edge density. 
Computations were carried out on a MacBook Air with an Apple M1 processor and 8~GB of RAM.  
The observed increase in computational cost aligns with the expected linear dependence on \(|E|\), 
confirming that the algorithm scales efficiently with network size and remains computationally tractable even for substantially larger graphs.}
\label{tab:complexity}
\end{table}

The speed results in Table~\ref{tab:complexity} are encouraging, especially given that computational efficiency was not a primary focus of this work.  
There remains substantial room for improvement–for instance, by porting the \texttt{R} code to \texttt{C++} or using more efficient operations. To conclude, Algorithm~\ref{alg:fullsampler} summarizes the steps described above that, together, constitute the conjugate single-site Gibbs with augmentation we employ to perform inference for our BT--SBM model.

\begin{algorithm}[htpb]
\begin{algorithmic}[1]
  \REQUIRE Initialization at $t=0$: $\{x_i^{(0)}=i\}_{i=1}^n$, $K^{(0)}=n$ (overdispersed), and $\lambda_k^{(0)} \sim \Gamma(a,b)$ independently.
  \FOR{$t = 0, 1, \ldots, T-1$}
    \STATE \textbf{Update $Z_{ij}$:} For each pair $(i,j)$ with $n_{ij}>0$, sample
    \[
        Z_{ij}^{(t+1)} \mid \bm\lambda^{(t)},\,\mathbf x^{(t)},\,\mathbf N \;\sim\; 
        \Gamma\!\Bigl(n_{ij},\, \lambda_{x_i^{(t)}}^{(t)} + \lambda_{x_j^{(t)}}^{(t)}\Bigr).
    \]
    \STATE \textbf{Update $\lambda_k$:} For each occupied cluster $k$, sample
    \[
        \lambda_k^{(t+1)} \mid \mathbf Z^{(t+1)},\,\mathbf x^{(t)},\,\mathbf W,\,\mathbf N 
        \;\sim\; \Gamma\!\Bigl(a + \textstyle\sum_{i\in I_k^{(t)}} w_i,\;\; b + \sum_{i\in I_k^{(t)}} Z_i^{(t+1)}\Bigr).
    \]
    \STATE \textbf{Update $x_i$:} For each $i=1,\dots,n$,
    \begin{enumerate}
      \item Remove $i$ from its current cluster; update $m_{-i,k}^{(t)}$. If that cluster becomes empty, remove it and re-index the clusters.
      \item Compute the unnormalized probabilities
      \[
         p\!\left(x_i^{(t+1)}=k \,\middle|\, \mathbf x_{-i},\, \bm\lambda^{(t+1)},\, \mathbf Z^{(t+1)},\, \mathbf W,\, \mathbf N \right),
         \quad k=1,\dots,K^{(t)}\!+1,
      \]
      according to~\eqref{eq:assignment_fc}.
      \item Normalize and sample a new value for $x_i$.
      \item If $x_i^{(t+1)} = K^{(t)}\!+1$, sample
      \[
        \lambda_{K^{(t)}+1}^{(t+1)} \mid \mathbf Z^{(t+1)},\, \mathbf W,\, \mathbf N 
        \;\sim\; \Gamma\!\Bigl(a + w_i,\;\; b + Z_i^{(t+1)}\Bigr),
      \]
      and set $K^{(t+1)} \leftarrow K^{(t)}\!+1$.
    \end{enumerate}
     \STATE \textbf{Global rescaling} To enforce the scale constraint $\prod_{k=1}^{K^{(t+1)}} \log \lambda_k^{(t+1)} = 0$, compute
\[
\log g^{(t+1)} \;=\; \frac{1}{K^{(t+1)}} \sum_{k=1}^{K^{(t+1)}} \log \lambda_k^{(t+1)}\]
and set 
$\log \lambda_k^{(t+1)} \leftarrow \log \lambda_k^{(t+1)} - \log g^{(t+1)}$
\;\; for all occupied $k$.
\ENDFOR
  \RETURN Collect $\{\,(\mathbf x^{(t)},\, \bm\lambda^{(t)},\, K^{(t)})\,\}_{t=1}^T$ from the posterior, discarding the first $B$ iterations.
\end{algorithmic}
\caption{Conjugate single-site Gibbs with augmentation for the BT--SBM\label{alg:fullsampler}}
\end{algorithm}

In the next section, we review methods to summarize the posterior samples; we then illustrate their performance of the BT-SBM in a simulation study, before finally turning to the real data application.

\section{Point Estimates and Uncertainty Quantification}
\label{sect:point_estimates}

Our MCMC sampler yields posterior samples of $\mathbf{x}$ and $\bm{\lambda}$, which we use to approximate the posterior distribution and derive summary statistics. After collecting the samples $\{\mathbf{x}^{(t)},\bm{\lambda}^{(t)}, K^{(t)}\}_{t=1}^T$, we discard the first $B$ iterations as burn-in to eliminate the non-stationary phase of the chain. Again, we emphasise, that $K^{(t)}$ is number of occupied clusters in $\mathbf{x}^{(t)}$ and hence it can be easily obtained from the latter. 

\subsection{Identifiability}\label{sect:identifiability}

The retained samples suffer from two well-known identifiability issues: \emph{label switching} and \emph{scale invariance}. We address both below.

\paragraph*{Label switching.}

The model likelihood is invariant under permutations of cluster labels. That is, for any permutation \(\pi\) of the labels,
\[
\mathcal{L}(\mathbf{W} \mid \mathbf{x}, \bm{\lambda}, \mathbf{N}) 
= 
\mathcal{L}(\mathbf{W} \mid \mathbf{x}_{\pi}, \bm{\lambda}_{\pi}, \mathbf{N}),
\]
where \(\mathbf{x}_\pi\) and \(\bm{\lambda}_\pi\) denote the permuted assignment vector and cluster strengths, respectively. As a result, the posterior distribution is also invariant under label permutations, and the MCMC sampler may visit the same partition structure under different labellings across iterations. This phenomenon is known as the \emph{label switching problem}, and it makes direct posterior summaries of \(\mathbf{x}\) or \(\bm{\lambda}\) meaningless unless the samples are relabelled consistently.

To address this, we relabel each sample by sorting the blocks in decreasing order of strength: the block with the largest $\lambda_k^{(t)}$ is assigned label 1, the second largest label 2, and so on. This post-processing aligns all samples to a common labelling convention. The procedure is detailed in Algorithm~\ref{alg:label_switch}.

\begin{algorithm}[htpb]
\caption{Label-switching correction \label{alg:label_switch}}
\begin{algorithmic}
\REQUIRE $\{\mathbf{x}^{(t)}, \bm{\lambda}^{(t)}\}_{t=B+1}^T$
\FOR{$t = 1,\ldots,T$}
  \STATE Extract the set of occupied clusters and their $\lambda_k^{(t)}$ values.
  \STATE Sort the $\lambda_k^{(t)}$ in decreasing order.
  \STATE Relabel clusters by rank in the sorted list (i.e., largest $\lambda_k^{(t)}$ becomes label 1, etc.).
\ENDFOR
\RETURN $\{\mathbf{x}_\pi^{(t)}, \bm{\lambda}_\pi^{(t)}\}_{t=B+1}^T$
\end{algorithmic}
\end{algorithm}

\paragraph*{Scale invariance.}

The BT--SBM likelihood is also invariant under multiplicative rescaling of the block strengths, and this issue is addressed above (see Sect.~\ref{subsec:hyperparam_choice}).

\subsection{Summarizing Posterior Samples}
\label{sec:point_estimates_uncertainty}

We summarize three key quantities from the posterior: 
(i) the partition $\mathbf{x}$, 
(ii) the cluster strengths $\bm{\lambda}$, 
and (iii) the number of occupied clusters $K$. Below, we outline how we extract point estimates and quantify posterior uncertainty for each.

\paragraph*{Partition point estimates.}

To summarize the posterior distribution over the $\{\mathbf{x}^{(t)}\}_{t=B+1}^T$ samples, we seek a consensus partition $\widehat{\mathbf{x}}$ that best represents the clustering structure. A principled and widely used approach is to minimize the expected \emph{Variation of Information (VI)} \citep{Meila2007}, an information-theoretic metric defined as:
\begin{equation}\label{eq:vi_distance}
\mathrm{VI}(A,B) = H(A) + H(B) - 2\,I(A,B),
\end{equation}
where $H(\cdot)$ is the Shannon entropy of a partition, and $I(A,B)$ is the mutual information between two partitions $A$ and $B$:
\begin{align}\label{eq:entropy_mi}
H(A) &= - \sum_{a} \frac{n_a}{n} \log \left( \frac{n_a}{n} \right), \qquad
I(A,B) = \sum_{a,b} \frac{m_{ab}}{n} \log \left( \frac{m_{ab} \cdot n}{n_a \cdot n_b} \right), 
\end{align}
where $n_a$ is the size of cluster $a$ under partition $A$, $n_b$ the size of cluster $b$ under $B$, and $m_{ab}$ the number of shared items between clusters $a$ and $b$.

Entropy is a measure of the uncertainty of a single partition, while mutual information captures the overlap between two. 
The Variation of Information (VI) is a metric on the space of partitions \citep{Meila2007} and has become a standard tool in Bayesian clustering \citep{Wade_2018, rastelli2018, Legramanti_2022}. 
The consensus partition $\widehat{\mathbf{x}}$ is defined as the minimizer of the posterior expected VI:
\[
\widehat{\mathbf{x}} 
= \arg\min_{\mathbf x} \mathbb{E}\bigl[\mathrm{VI}(\mathbf x, \mathbf{x}^{(t)}_\pi)\bigr]
\approx 
\arg\min_{x} \frac{1}{T-B} \sum_{t=B+1}^{T} \mathrm{VI}(\mathbf x, \mathbf{x}_\pi^{(t)}).
\]

where each sample $\mathbf{x}^{(t)}_\pi$ is relabelled to account for label switching, as described above. This estimator, efficiently implemented in \texttt{R} through the \texttt{mcclust} and \texttt{mcclust.ext} packages \citep{Wade_2018}, tends to penalize the formation of small clusters, yielding a parsimonious summary partition.

A partition estimate summarises the full posterior sample $\{\mathbf{x}^{(t)}\}_{t = B + 1}^T$ into a single labelling. To fully exploit the richness of the posterior, it is therefore crucial to assess and appropriately represent the \emph{uncertainty} surrounding $\widehat{\mathbf{x}}$. A common approach is to compute the co-clustering probability between pairs of nodes and represent it as a heatmap, i.e., the similarity or co-clustering matrix \citep{Legramanti_2022}. However, such heatmaps can be difficult to interpret and, as noted by \citet{Wade_2018}, they tend to understate posterior uncertainty.  

An appealing alternative proposed by \citet{Wade_2018} is the construction of a \emph{credible ball}–the analogue of a credible interval in the high-dimensional space of partitions–which captures the range of plausible clusterings around the point estimate $\widehat{\mathbf{x}}$. Partitions on the edge (or surface) of this ball provide an intuitive sense of alternative groupings that remain well supported by the posterior distribution.
 
Define an $\varepsilon-$ball around $\widehat{\mathbf{x}}$ of size $\varepsilon$ as 
\[
B_{\varepsilon}(\widehat{\mathbf{x}}) := 
\big\{
\mathbf{x} : \mathrm{VI}(\mathbf{x}, \widehat{\mathbf{x}}) \le \varepsilon
\big\}. 
\]
Then the posterior probability of $B_{\varepsilon}(\widehat{\mathbf{x}})$ can be expressed as  
\[
P\big(B_\varepsilon(\widehat{\mathbf{x}}) \mid \text{data}\big)
\;=\;
E\big[ \mathbb{I}\{ \mathrm{VI}(\mathbf{x},\widehat{\mathbf{x}}) \le \varepsilon \} \mid \text{data} \big]
\;\approx\;
\frac{1}{T - B} \sum_{t = B + 1}^{T} 
\mathbb{I}\big\{ \mathrm{VI}(\mathbf{x}^{(t)}, \widehat{\mathbf{x}}) \le \varepsilon \big\},
\]
where $\mathbb{I}\{\mathrm{VI}(\mathbf{x}^{(t)},\widehat{\mathbf{x}}) \le \varepsilon\} = 1$ if $\mathbf{x}^{(t)}$ lies within a $\varepsilon$ VI-distance from $\widehat{\mathbf{x}}$.  
The $(1 - \alpha)$ \emph{credible ball} around $\widehat{\mathbf{x}}$ is then defined as the smallest metric ball, $B_{\varepsilon^*}(\widehat{\mathbf{x}})$, where  
\[
\varepsilon^* = 
\inf \big\{ \varepsilon \ge 0 : 
P\big(B_\varepsilon(\widehat{\mathbf{x}}) \mid \text{data}\big) \ge 1 - \alpha
\big\}.
\]
In our application, we set $\alpha = 0.05$, yielding a 95\% credible ball. Intuitively, this region contains the set of partitions that are \emph{sufficiently close} to the optimal partition $\widehat{\mathbf{x}}$ under the VI metric to collectively account for 95\% of the posterior probability. A small $\varepsilon^*$ indicates a concentrated posterior, reflecting strong confidence in the estimated clustering.  

Once the credible ball is constructed, it is often insightful to examine its \emph{surface}, that is, the set of partitions $\mathbf{x}$ such that $\mathrm{VI}(\mathbf{x}, \widehat{\mathbf{x}}) = \varepsilon^*$.  
Among these, \citet{Wade_2018} identify three representative types:
\begin{itemize}
    \item \textbf{Vertical upper bounds}: surface partitions with the \emph{fewest} clusters, corresponding to the coarsest partition within the 95\% credible ball and maximally distant from $\widehat{\mathbf{x}}$;
    \item \textbf{Vertical lower bounds}: surface partitions with the \emph{largest} number of clusters, corresponding to the finest partition within the 95\% credible ball and maximally distant from $\widehat{\mathbf{x}}$;
    \item \textbf{Horizontal bounds}: surface partitions that are maximally distant from $\widehat{\mathbf{x}}$, irrespective of the number of clusters.  
\end{itemize}

Inspecting these surface partitions provides an intuitive view of how posterior uncertainty distributes across alternative but still plausible clusterings.  
The computation of credible balls and their corresponding bounds is implemented in the \texttt{R} package \texttt{mcclust} \citep{Wade_2018}.

\paragraph*{Number of clusters \texorpdfstring{$K$}{K} and model complexity}

The credible ball can also be used to quantify the uncertainty surrounding the number of clusters $K$. Specifically, we define $\widehat{K}^{(\text{VI})}$ as the number of groups in the partition point estimate $\widehat{\mathbf{x}}$ obtained under the VI loss, and denote by $K_{\mathrm{ub}}$ and $K_{\mathrm{lb}}$ the numbers of clusters associated with the \emph{vertical upper} (coarsest) and \emph{vertical lower} (finest) bounds, respectively \citep{Wade_2018}.  
The resulting range,
\[
\widehat{K}^{(\text{VI})}_{[K_{\mathrm{ub}},\,K_{\mathrm{lb}}]},
\]
provides a compact measure of the variability of $K$ within the 95\% credible region around $\widehat{\mathbf{x}}$.  

These quantities complement the direct inspection of the number of occupied clusters $K$ across MCMC samples. We approximate its posterior distribution by collecting $\{K^{(t)} = K(\mathbf{x}^{(t)})\}_{t = B + 1}^T$ and summarizing its empirical frequencies. As a point estimate, we report the posterior mode
\[
\widehat{K} \coloneqq \mathrm{mode}\{K^{(t)}\},
\]
interpreted as the most probable number of clusters supported by the posterior under the Gnedin prior, together with its 95\% credible interval.

These two estimators of $K$, $(\widehat{K}$ and $\widehat{K}^{(\text{VI})})$, along with their associated uncertainty measures, are not theoretically guaranteed to coincide. In practice, however, we observe substantial agreement between them across most seasons, with occasional discrepancies reflecting the distinct nature of the two summaries.

\paragraph*{Strength estimates.}

Given the normalized and relabelled MCMC samples \(\bm{\lambda}_\pi^{(t)}\), we summarize the strength of each block \(k\) by its posterior mean:
\[
\widehat{\lambda}_k = \frac{1}{T - B} \sum_{t = B+1}^{T} \lambda_{\pi,k}^{(t)}, \quad \text{for } k=1,\dots,K.
\]

\section{Analysis of the last 23 years of men's tennis}

We now present a detailed analysis of the 2017/2018 ATP men's tennis season, highlighting the main outputs of our model. We focus on this season because of the substantial uncertainty regarding the number of clusters. 
This case study also serves to illustrate the BT--SBM more concretely in the context of real competition data. 

We subsequently repeat the analysis across all seasons from 2000/2001 to 2022/2023 (23 seasons in total), a period spanning the rise, peak, and eventual decline of the dominance exerted by Federer, Nadal, Djokovic, and Murray---the quartet commonly referred to as the ``Big Four.'' 

This era profoundly shaped men's tennis, and our model captures its evolution through changes in the inferred competitive structure over time.
For each season, our MCMC algorithm (described in Sect.~\ref{sec:posterior_inference}) was independently run for 30{,}000 iterations, discarding the first 10{,}000 as burn-in.  
Fitting the model to all 23 seasons under these specifications required approximately 35 minutes in total.  
We used a Gnedin prior for $\mathbf{x}$ with hyperparameter $\gamma = 0.8$, and a Gamma prior for the rate parameters with shape $a = 2$ and rate $b = \exp\{\psi(a)\} = 1.526$.

The label switching correction algorithm (Alg.~\ref{alg:label_switch}) was subsequently applied. All computations, regarding both the simulation study and the application to the real data, were implemented in \texttt{R}, and the associated code and data are available at: 
\href{https://github.com/laposanti/BT-SBM-Bradley-Terry-Stochastic-Block-Model}{https://github.com/laposanti/BT-SBM}.

\subsection{The 2017/2018 ATP men's Season}

We examine in detail the 2017/2018 ATP season. 
Table~\ref{tab:single_season_prob_clust} reports the posterior probabilities for different numbers of blocks under the Gnedin-type prior. 
The posterior slightly favours a four-block model ($\widehat{K}=4$), although the 95\% credible interval on $K$ spans configurations between 3 and 7 clusters.

\begin{table}[htpb]
  \centering
  \begin{tabular}{lrrrrrrrr}
\toprule
  Season & 2 & 3 & 4 & 5 & 6&7&8 \\
  \midrule
2017/2018 & -- & 0.259& \cellcolor{clayOrange} 0.315& 0.216& 0.110 &0.056& 0.025 \\
  \bottomrule
  \end{tabular}
  \caption{
Posterior distribution of the number of clusters (\(K\)) inferred for the 2017/2018 ATP men's season under the Gnedin-type prior. 
Each entry reports the posterior probability of a given \(K\), with the most probable configuration (\(K=4\)) highlighted in orange. 
Although the four-block model achieves the highest support (\(p(K=4 \mid \mathbf{W}) = 0.315\)), the posterior mass remains substantial across \(K \in \{3,\dots,7\}\), indicating non-negligible uncertainty about the true number of latent groups.
}
\label{tab:single_season_prob_clust}
\end{table}

\begin{figure}[htpb]
  \centering

  \includegraphics[width=\linewidth]{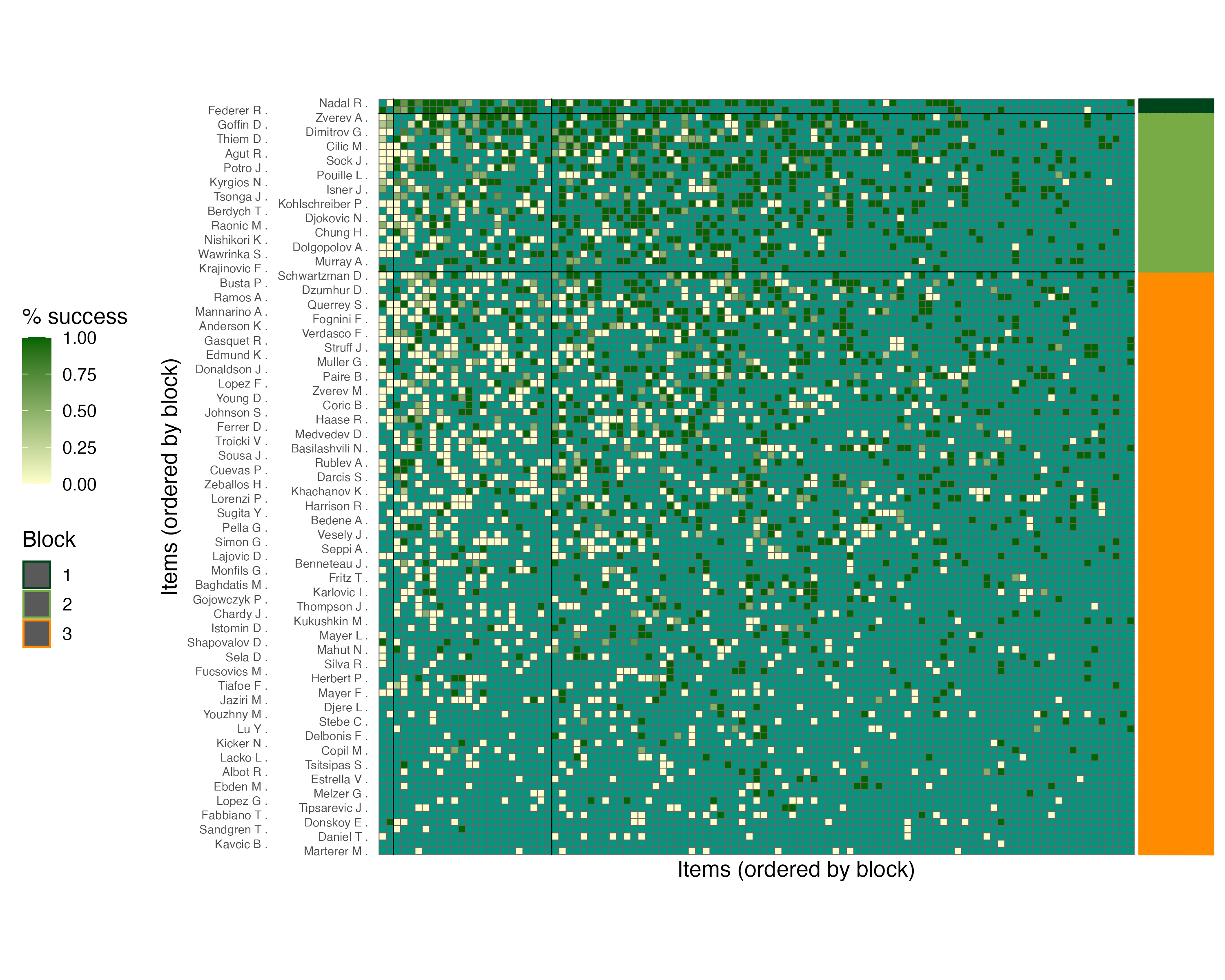}
 \caption{Reordered adjacency matrix of ATP matches during the 2017/2018 season, obtained from the estimated partition $\mathbf{\hat{x}}$ with \(K=3\). 
  Rows and columns correspond to players, ordered first by inferred cluster membership (from 1 = strongest to 3 = weakest) and then by their marginal win proportions. 
  Coloured bars on the left mark the three inferred clusters (dark-green, light green, and orange). 
  Each pixel represents the empirical proportion of wins of the row player against the column player, with a white-to-green gradient indicating increasing success rates for the row player.
  The uniform colour pattern along the main diagonal reveals strong within-cluster balance. 
  In contrast, the top-right corner is dominated by darker shades, corresponding to encounters between top-block and weakest-block players, where stronger opponents prevail almost systematically, as expected. 
  Conversely, the bottom-left corner is mostly white, reflecting matches in which the weakest blocks concede to the elite group.
  Players’ names are staggered across two columns to improve readability. 
  Nadal corresponds to the first row, Federer to the second, Zverev to the third, and so on.  \label{fig:adjacency-reordered} }
\end{figure}

Given this diffuse posterior, we next identify a representative point estimate. 
We obtain an estimated partition $\widehat{\mathbf{x}}$ via VI loss (Sect.~\ref{sec:point_estimates_uncertainty}) and reorder the data accordingly, as displayed in Figure~\ref{fig:adjacency-reordered}. 
Because the VI loss tends to penalise small clusters, it produces a parsimonious summary (\(\widehat{K}^{(\text{VI})} =3 < \widehat{K} = 4\)). 
In this case, the VI point estimate with \(\widehat{K}^{(\text{VI})} = 3_{[3,11]}\) reveals a coarse but interpretable block structure that divides players into three main groups: a narrow elite, a compact middle section, and a broad lower tier of weaker competitors.

The top group isolates \emph{Nadal} and \emph{Federer}, reflecting the marked competitive gap of the 2017–2018 season. 
This dominance is clearly visible in the top-right corner of the reordered adjacency matrix, where dark-green pixels correspond to head-to-head winning probabilities close to one.  
Moving across the matrix, the bottom-left corner tells the complementary story: its prevalence of white pixels indicates that players in the weakest block rarely beat those above them.  

Along the main diagonal, the three squared blocks correspond to within-block matchups.  
These regions are almost uniform in colour, consistent with the assumption that for players belonging to the same block the probability of victory is roughly $0.5$, a consequence of sharing the same strength parameter (\(\hat{\lambda}_i = \hat{\lambda}_j \iff x_i = x_j\)).   Taken together, these visual patterns reveal a sharply hierarchical yet clustered structure: a small elite dominates, a compact group of contenders follows closely, and a large set of statistically similar players forms the broad lower tier.

\begin{figure}[htpb]
    \centering
    \includegraphics[width=\linewidth]{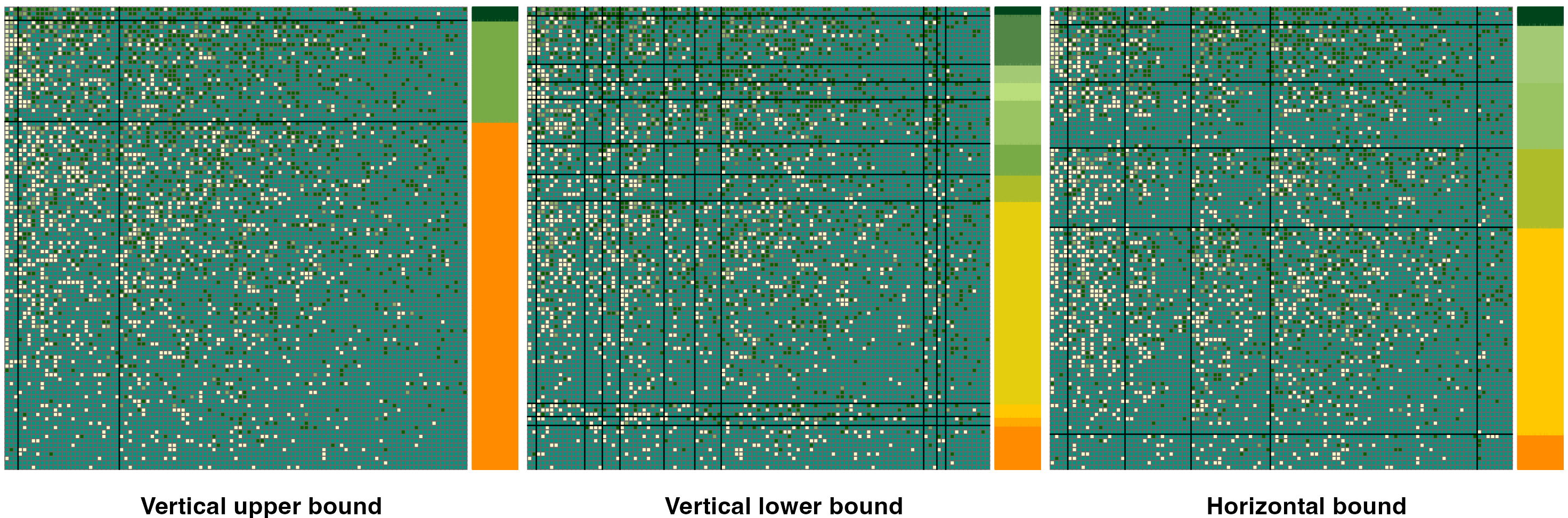}
   \caption{Reordered adjacency matrix of the 2017–2018 ATP season. From left to right, the matrix is reordered according to the vertical upper bound, vertical lower bound, and horizontal bound partitions. Magnified views of these plots are shown in Appendix~\ref{appendix:edge-partitions}. A detailed guide to interpreting this type of graph is provided in Fig.~\ref{fig:adjacency-reordered}.}
    \label{fig:placeholder}
\end{figure}

Although the VI estimate provides a concise summary, the posterior retains non-negligible uncertainty, supporting partitions up to \(K = 11\) within the 95\% credible ball.  
The \emph{vertical upper}, \emph{vertical lower}, and \emph{horizontal} bounds partitions are reported in Fig.~\ref{fig:placeholder}, which illustrates how uncertainty propagates across the hierarchy.\footnote{Magnified versions of these configurations are provided in Appendix~\ref{appendix:edge-partitions}.}
The coarsest partition (vertical upper bound) nearly coincides with the VI estimate, confirming that the VI criterion favours fewer and broader blocks.  
At the opposite extreme, the finest configuration (vertical lower bound) supports higher resolutions: the middle and weakest tiers split into six and four sub-blocks, respectively.  
Between these extremes, the horizontal bound provides an intermediate \(K = 6\) solution with more balanced block sizes.  

Compared with the \(K = 3\) configuration, these higher-resolution partitions introduce new blocks primarily in two regions: between block~2 and block~3, and at the bottom of block~3.  
The elite group remains unchanged, while the middle and lower parts of the hierarchy are refined by the addition of further layers.  
This indicates that the global order of strengths is stable, whereas the granularity of the lower tiers remains more flexible.  
Depending on the chosen resolution, these additional groups can either be kept separate or merged into broader tiers, yielding more parsimonious summaries.  
In this sense, the credible ball spans a continuum of plausible resolutions–from coarse to fine–rather than representing competing hierarchies of the data.  

At the limit, the standard Bradley–Terry model corresponds to the extreme case in which every player forms their own block–a strict global order that enforces sharp distinctions even where the data provide little evidence for them.  
Our block-based approach, by contrast, preserves the stable clustered structure while explicitly accounting for the uncertainty arising from group separation.

\paragraph*{Uncertainty in cluster assignments}

\begin{figure}[htpb]
  \centering
  \includegraphics[width=\linewidth]{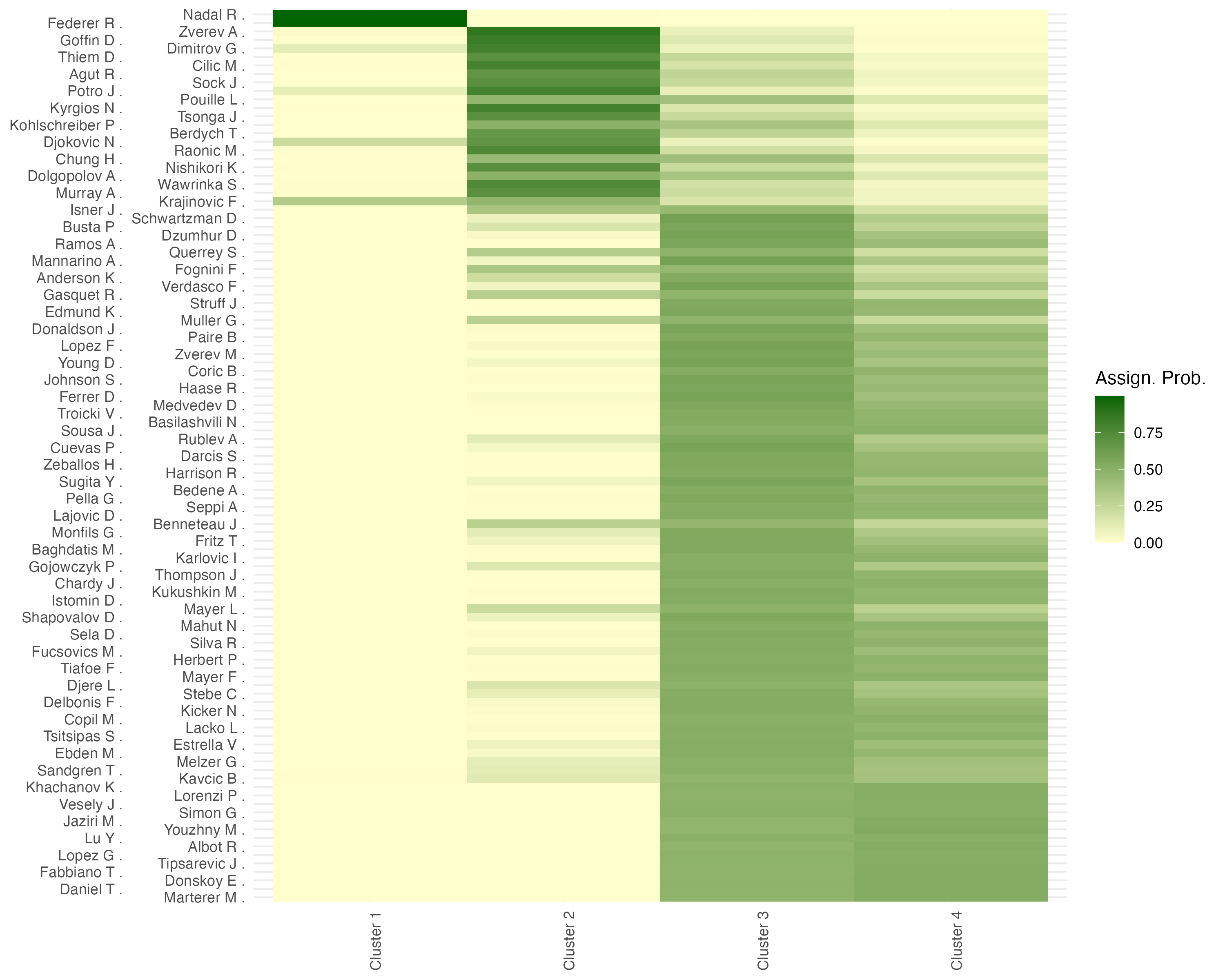}
  \caption{Posterior probabilities of cluster membership \(p(x_i = k \mid \mathbf{W})\) for each player in the 2017/2018 ATP season, conditioned on MCMC draws with \(K=4\).  
Each row corresponds to a player and each column to an inferred block label (14).  
Players are ordered as in Figure~\ref{fig:adjacency-reordered}, and darker shades indicate higher posterior probabilities.  
Players at the top and bottom of the hierarchy show near-certain assignments, whereas uncertainty concentrates between blocks 2 and 3 and at the bottom of block 3, where several players display diffuse membership--compatible also with block 4.  
This pattern suggests that a subset of these players could plausibly form an additional, well-supported fourth block, consistent with the posterior evidence on \(K\).  
Names are staggered across two columns for readability.\label{fig:ass-prob}}
\end{figure}

We can examine this uncertainty more directly by looking at the posterior probabilities of cluster membership–that is, how confidently each player is assigned to a block.  
Figure~\ref{fig:ass-prob} displays \(p(x_i = k \mid \mathbf{W})\), restricted to draws with \(K^{(t)} = 4\) to isolate allocation uncertainty from uncertainty in \(K\) and to illustrate how a fourth block could also be supported.  
Players are ordered as in Figure~\ref{fig:adjacency-reordered} to facilitate direct comparison with the point-estimate adjacency matrix above.

Two main factors drive uncertainty in cluster allocation: limited data and discontinuous performance.  
The first mainly affects players at the bottom of the hierarchy, for whom we observe few matches; as data become sparse, posterior assignments become more diffuse.  
In contrast, the top tier–where data are abundant–shows highly concentrated memberships (\(p(x_i = k \mid \mathbf{W}) \approx 1\)).  

For players between blocks 1 and 2–where the network is dense–uncertainty instead arises from irregular or exceptional seasons.  
Examples include fast-rising competitors such as \emph{Filip Krajinović}, whose 2017 Paris Masters final marked a sudden ascent, and established players like \emph{Novak Djokovic} and \emph{Juan Martín del Potro}, whose injury-interrupted seasons blur their position within the hierarchy.

Finally, allocation uncertainty is highest between blocks 2 and 3 and at the bottom of block 3, where players are hardest to separate with confidence.  
These players could plausibly form an additional, distinct group, reinforcing the patterns observed in the vertical lower and horizontal partitions described above.

\paragraph*{Player-level strength uncertainty}

\begin{figure}[htpb]
  \centering
  \includegraphics[width=0.85\linewidth]{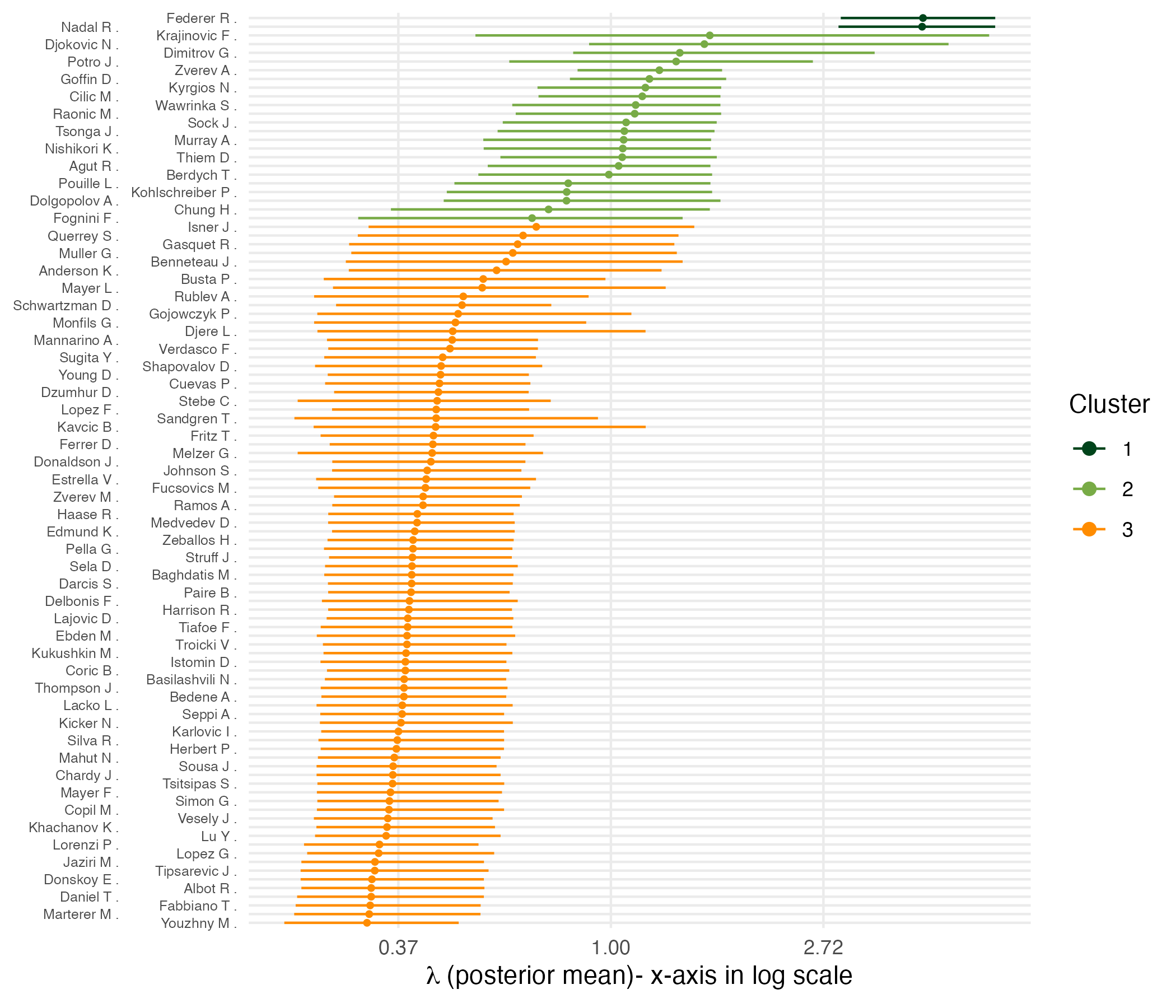}
  \caption{
Posterior estimates of player-specific strength parameters \(\tilde{\lambda}_i\) for the 2017/2018 ATP season, conditioned on MCMC draws with \(K^{(t)}=3\). 
Players are ordered by their posterior mean strength, with 95\% HPD credible intervals. 
Elite players are sharply separated, whereas mid-ranked competitors exhibit wide overlaps, mirroring the assignment uncertainty in Figure~\ref{fig:ass-prob}. 
This alignment underscores how the posteriors uncertainty concentrates in the middle region of the hierarchy and at the the borders between blocks.
  \label{fig:lambda_uncertainty}}
\end{figure}

The uncertainty visible in cluster assignments is mirrored in the posterior distribution of player-specific strengths \(\tilde{\lambda}_i\). 
Figure~\ref{fig:lambda_uncertainty} reports posterior means and 95\% HPD intervals from relabelled draws (Algorithm~\ref{alg:label_switch}), with
\begin{equation}\label{eq:lambda_mapping}
  \tilde{\lambda}_i^{(t)} \coloneqq \lambda_{\pi, x_i^{(t)}}^{(t)} \quad \text{for } i \in 1,\ldots,n,
\end{equation}
filtered to \(K^{(t)}=3\). 
Elite players (Nadal, Federer, Zverev) display large \(\tilde{\lambda}_i\) values and tight credible intervals; the weakest block is equally well separated at the lower end. 
In contrast, mid-tier players exhibit broad and overlapping HPD intervals–precisely where the posterior mass is most diffuse and clustering least stable.  

Taken together, the credible-ball partitions (Figs.~\ref{fig:placeholder}), the uncertainty analyses in Figures~\ref{fig:ass-prob} and~\ref{fig:lambda_uncertainty} highlight a key aspect of the posterior structure. 
A stable elite group is followed by a more unstable block of players that can split into smaller sub-groups depending on the resolution of the partition. 
In this middle region, block assignments and strength estimates are inherently harder to pin down, and such uncertainty is well-accounted in the BT-SBM, which is robust to the variability of tennis competition.  In this sense, the block-based representation offers a more faithful summary: it preserves the clear dominance relations at the top and bottom while acknowledging the ambiguity that naturally arises in the middle of the hierarchy.

\subsection{\texorpdfstring{BT--SBM analysis: all seasons $2000/01$--$2022/23$}{BT--SBM analysis: all seasons 2000/01--2022/23}}
\label{sec:big_four}

We now apply our model independently across seasons from the early 2000s to the early 2020s–a period spanning the rise, dominance, and gradual decline of the so-called \emph{Big~Four}.
This era is marked by a remarkably stable elite group of players that captured the lion’s share of major titles, profoundly shaping the competitive structure of the men's tour. 
In the seasons immediately preceding the COVID-19 disruption, we observe the fading of this dominance and the emergence of a new, more fragmented cluster of elite contenders.

\subsubsection{Model comparison: BT--SBM vs. standard BT}\label{sec:model_comparison}

Before exploring the temporal evolution of player clusters, we first ask whether the BT--SBM provides a tangible improvement over the standard Bradley--Terry (BT) model in terms of out-of-sample predictive accuracy and model fit.

In the standard BT model, each player \(i\) has an individual strength \(\lambda_{i}>0\). 
With independent priors \(\lambda_{i} \sim \Gamma(a,b)\) and the identifiability constraint \(\prod_i \lambda_{i} = 1\), the posterior is
\[
p_{\text{BT}}(\bm\lambda \mid \mathbf W)
\;\propto\;
\Biggl[\prod_{i=1}^n \lambda_{i}^{\,a-1} e^{-b\lambda_{i}}\Biggr]
\;\times\;
\prod_{1\le i<j\le n}
\biggl(\frac{\lambda_{i}}{\lambda_{i}+\lambda_{j}}\biggr)^{w_{ij}}
\biggl(\frac{\lambda_{j}}{\lambda_{i}+\lambda_{j}}\biggr)^{w_{ji}},
\qquad \text{subject to } \prod_{i=1}^n \lambda_{i}=1.
\]
Regarding the BT-SBM, its corresponding posterior $p(\bm\lambda_{\text{BT-SBM}}, \mathbf x \mid \mathbf W)$ is defined in \eqref{eq:full_posterior_detailed}.

To assess predictive performance across the two models $\mathcal{M}\in\{\text{BT},\text{BT-SBM}\}$, we perform leave-one-out cross-validation using Pareto-smoothed importance sampling (PSIS--LOO) \citep{vehtariPracticalBayesianModel2016}.

For each directed edge \((i, j) \in E\) with at least one recorded match, we evaluate the model’s ability to predict the outcome \(w_{ij}\) when it is held out from the data. 
That is, for each edge, model $\mathcal{M}$ is effectively fit on \(\mathbf{W}_{-ij}\)---the data matrix excluding \(w_{ij}\)---and predictive accuracy is assessed via the log posterior predictive density:
\[
 \operatorname{lpd}_{ij,\mathcal{M}} 
 = \log\bigl[\,p_\mathcal{M}\bigl(w_{ij} \mid \mathbf{W}_{-ij}\bigr)\,\bigr].
\]
In practice, PSIS--LOO approximates this by reweighting posterior samples 
\(\Theta_\mathcal{M}^{(t)} \sim p_\mathcal{M}(\Theta_\mathcal{M} \mid \mathbf{W})\)
using importance weights that account for the omission of \(w_{ij}\), yielding
\[
 \widehat{\operatorname{lpd}}_{ij,\mathcal{M}} 
 = \sum_{t=B+1}^{T} r_{ij,\mathcal{M}}^{(t)} 
   \log\bigl[\,p_\mathcal{M}\bigl(w_{ij} \mid \Theta_\mathcal{M}^{(t)}\bigr)\,\bigr],
\]
where \(r_{ij,\mathcal{M}}^{(t)}\) are normalised PSIS importance weights.

Summing over all relevant edges yields the expected log pointwise predictive density (ELPD) for model $\mathcal{M}$,
\[
 \widehat{\text{ELPD}}_{\mathcal{M}} 
 = \sum_{(i,j)\in E} \widehat{\operatorname{lpd}}_{ij,\mathcal{M}},
\]
which serves as a measure of out-of-sample predictive performance. 
Higher ELPD values indicate better generalisation and model fit. 
We computed $\widehat{\text{ELPD}}_{\mathcal{M}}$ for both models in each of the 23 seasons (2000--2022) and summarised their difference as
\[
 \Delta\text{ELPD} 
  = \widehat{\text{ELPD}}_{\text{BT-SBM}} 
  - \widehat{\text{ELPD}}_{\text{BT}},
\]
with one--standard--error bounds 
\(
\text{SE}_\Delta 
  = \sqrt{\text{SE}_{\text{BT-SBM}}^2+\text{SE}_{\text{BT}}^2}/2
\)
obtained from the PSIS diagnostics.

\begin{figure}[htpb]
  \centering
  \includegraphics[width=\linewidth]{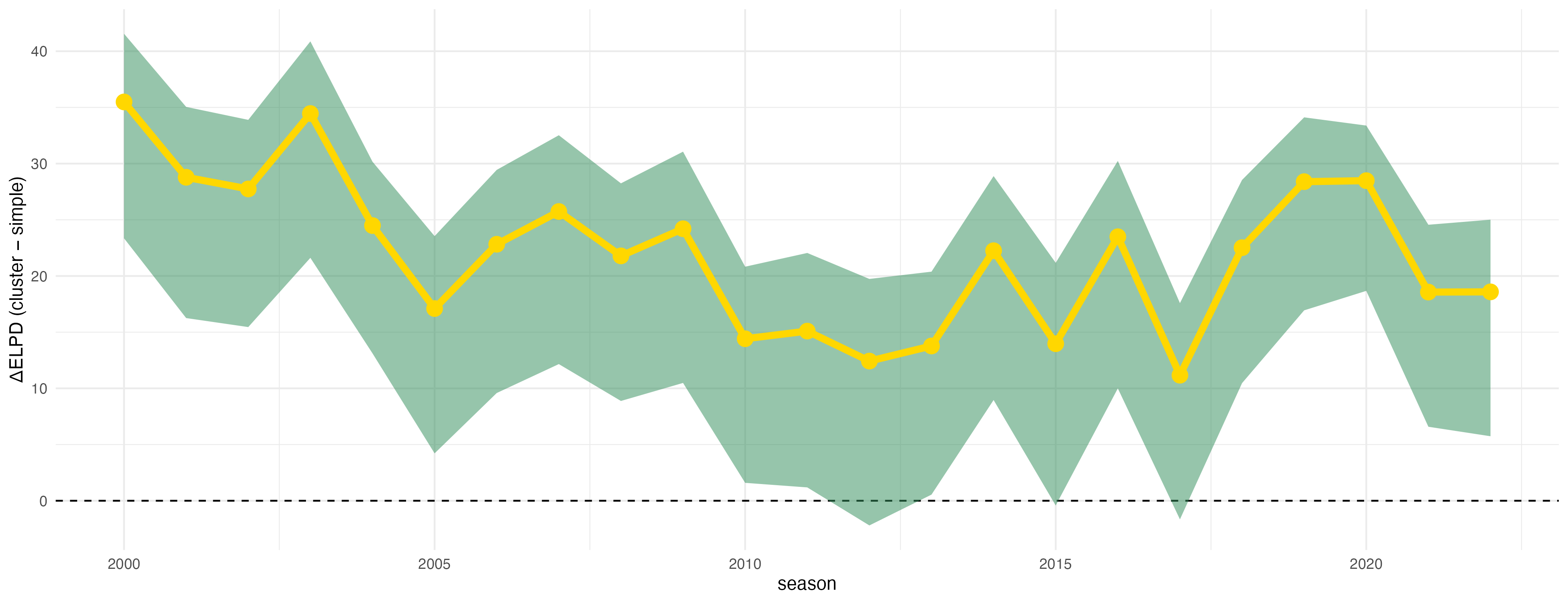}
  \caption{Predictive gain of the clustered BT--SBM over the classical BT model for each season, expressed as the difference in expected log predictive density (\(\Delta \mathrm{ELPD}\)). 
  The yellow line shows the mean \(\Delta \mathrm{ELPD}\) across seasons, with shaded bands denoting \(\pm 1\,\mathrm{SE}_{\Delta}\) obtained from PSIS--LOO. 
  Positive values indicate superior predictive performance of the BT--SBM relative to the standard BT model. 
  The consistently positive trend highlights the improved fit achieved by incorporating block structure into the model. 
  \label{fig:delta_elpd}}
\end{figure}

The clustered BT--SBM outperformed the classical BT in every season. 
The median gain was $22.99$ ELPD points (median $\text{SE}_\Delta$~=~12.1), and in approximately $87\%$ of seasons the improvement exceeded one SE, providing moderate to strong evidence in favour of the clustered specification (see Fig.~\ref{fig:delta_elpd}).

\begin{table}[htpb]
  \centering
  \begin{tabular}{lcccc}
    \toprule
    Statistic & Min & Median & Mean & Max \\
    \midrule
    $\Delta \mathrm{ELPD}$ & 11.17 & 22.52 & 21.99 & 35.49 \\
    \bottomrule
  \end{tabular}
  \vspace{1mm}

  \noindent
  Proportion of seasons with $\Delta \mathrm{ELPD} > \mathrm{SE}_{\Delta}$: 0.87
  
  \caption{Season-aggregated model comparison metrics. 
  The table reports summary statistics of the predictive gain (\(\Delta \mathrm{ELPD}\)) of the clustered BT--SBM over the classical BT model across all seasons. 
  Values above zero indicate that the clustered BT--SBM provides a better out-of-sample predictive fit. 
  In 87\% of the seasons, the predictive improvement exceeded one standard error, further supporting the robustness of the clustering formulation. 
  \label{tab:elpd_summary}}
\end{table}

\subsubsection{Results' analysis}

We now turn to the core output of our longitudinal analysis, which examines how the inferred cluster structure evolves across seasons. Table~\ref{tab:Kposterior_across_seasons} reports the posterior probabilities associated with different values of \(K\), with the modal number of blocks for each year highlighted using a consistent colour palette adopted throughout this section\footnote{The colour scale has been manually chosen: a green tone, reminiscent of grass courts, denotes \(K=3\), whereas an orange shade, akin to the clay tournaments, represents \(K=4\).}.
A few broad patterns stand out in the evolution of cluster structure over time.

\begin{table}[htpb]
\centering
\begin{tabular}{lrrrrrrr}
\toprule
Season $\backslash \widehat{K}$ & 2 & 3 & 4 & 5 & 6 & 7 & 8\\
\midrule
\cellcolor{gray!10}{2000/2001} & \cellcolor{gray!10}{0.120} & \cellcolor{wimbledonGreen}{0.510} & \cellcolor{gray!10}{0.243} & \cellcolor{gray!10}{0.078} & \cellcolor{gray!10}{0.030} & \cellcolor{gray!10}{0.011} & \cellcolor{gray!10}{0.005}\\
2001/2002 & 0.013 & \cellcolor{wimbledonGreen}{0.565} & 0.269 & 0.100 & 0.035 & 0.013 & 0.003\\
\cellcolor{gray!10}{2002/2003} & \cellcolor{gray!10}{0.335} & \cellcolor{wimbledonGreen}{0.370} & \cellcolor{gray!10}{0.156} & \cellcolor{gray!10}{0.076} & \cellcolor{gray!10}{0.037} & \cellcolor{gray!10}{0.014} & \cellcolor{gray!10}{0.007}\\
2003/2004 & -- & \cellcolor{wimbledonGreen}{0.520} & 0.303 & 0.115 & 0.039 & 0.014 & 0.006\\
\cellcolor{gray!10}{2004/2005} & \cellcolor{gray!10}{0.002} & \cellcolor{wimbledonGreen}{0.512} & \cellcolor{gray!10}{0.288} & \cellcolor{gray!10}{0.122} & \cellcolor{gray!10}{0.048} & \cellcolor{gray!10}{0.018} & \cellcolor{gray!10}{0.007}\\
2005/2006 & -- & 0.218 & \cellcolor{clayOrange}{0.379} & 0.241 & 0.097 & 0.041 & 0.015\\
\cellcolor{gray!10}{2006/2007} & \cellcolor{gray!10}{--} & \cellcolor{gray!10}{0.017} & \cellcolor{clayOrange}{0.299} & \cellcolor{gray!10}{0.285} & \cellcolor{gray!10}{0.179} & \cellcolor{gray!10}{0.107} & \cellcolor{gray!10}{0.061}\\
2007/2008 & -- & \cellcolor{wimbledonGreen}{0.534} & 0.285 & 0.113 & 0.040 & 0.016 & 0.006\\
\cellcolor{gray!10}{2008/2009} & \cellcolor{gray!10}{--} & \cellcolor{gray!10}{0.172} & \cellcolor{clayOrange}{0.372} & \cellcolor{gray!10}{0.238} & \cellcolor{gray!10}{0.120} & \cellcolor{gray!10}{0.057} & \cellcolor{gray!10}{0.026}\\
2009/2010 & -- & 0.051 & \cellcolor{clayOrange}{0.433} & 0.300 & 0.124 & 0.055 & 0.022\\
\cellcolor{gray!10}{2010/2011} & \cellcolor{gray!10}{--} & \cellcolor{gray!10}{0.278} & \cellcolor{clayOrange}{0.322} & \cellcolor{gray!10}{0.201} & \cellcolor{gray!10}{0.103} & \cellcolor{gray!10}{0.051} & \cellcolor{gray!10}{0.024}\\
2011/2012 & -- & 0.151 & \cellcolor{clayOrange}{0.382} & 0.252 & 0.120 & 0.055 & 0.024\\
\cellcolor{gray!10}{2012/2013} & \cellcolor{gray!10}{--} & \cellcolor{gray!10}{0.023} & \cellcolor{clayOrange}{0.312} & \cellcolor{gray!10}{0.289} & \cellcolor{gray!10}{0.184} & \cellcolor{gray!10}{0.105} & \cellcolor{gray!10}{0.048}\\
2013/2014 & -- & 0.021 & \cellcolor{clayOrange}{0.365} & 0.303 & 0.176 & 0.080 & 0.035\\
\cellcolor{gray!10}{2014/2015} & \cellcolor{gray!10}{--} & \cellcolor{gray!10}{0.024} & \cellcolor{clayOrange}{0.418} & \cellcolor{gray!10}{0.305} & \cellcolor{gray!10}{0.160} & \cellcolor{gray!10}{0.061} & \cellcolor{gray!10}{0.023}\\
2015/2016 & -- & 0.021 & \cellcolor{clayOrange}{0.504} & 0.282 & 0.116 & 0.049 & 0.019\\
\cellcolor{gray!10}{2016/2017} & \cellcolor{gray!10}{--} & \cellcolor{gray!10}{0.082} & \cellcolor{clayOrange}{0.396} & \cellcolor{gray!10}{0.278} & \cellcolor{gray!10}{0.133} & \cellcolor{gray!10}{0.066} & \cellcolor{gray!10}{0.029}\\
2017/2018 & -- & 0.259 & \cellcolor{clayOrange}{0.315} & 0.216 & 0.110 & 0.056 & 0.025\\
\cellcolor{gray!10}{2018/2019} & \cellcolor{gray!10}{--} & \cellcolor{wimbledonGreen}{0.346} & \cellcolor{gray!10}{0.332} & \cellcolor{gray!10}{0.175} & \cellcolor{gray!10}{0.086} & \cellcolor{gray!10}{0.035} & \cellcolor{gray!10}{0.016}\\
2019/2020 & 0.010 & \cellcolor{wimbledonGreen}{0.501} & 0.299 & 0.122 & 0.044 & 0.016 & 0.005\\
\cellcolor{gray!10}{2020/2021} & \cellcolor{gray!10}{0.352} & \cellcolor{wimbledonGreen}{0.371} & \cellcolor{gray!10}{0.165} & \cellcolor{gray!10}{0.072} & \cellcolor{gray!10}{0.024} & \cellcolor{gray!10}{0.009} & \cellcolor{gray!10}{0.005}\\
2021/2022 & -- & 0.248 & \cellcolor{clayOrange}{0.341} & 0.210 & 0.110 & 0.054 & 0.021\\
\cellcolor{gray!10}{2022/2023} & \cellcolor{gray!10}{0.003} & \cellcolor{wimbledonGreen}{0.387} & \cellcolor{gray!10}{0.314} & \cellcolor{gray!10}{0.156} & \cellcolor{gray!10}{0.077} & \cellcolor{gray!10}{0.038} & \cellcolor{gray!10}{0.014}\\
\bottomrule
\end{tabular}
\caption{
Posterior probabilities of the number of clusters (\(K\)) across ATP men's seasons (2000/2001--2022/2023), inferred under the Gnedin-type prior. 
Each row corresponds to a season, and each column to a candidate value of \(K\). 
The modal number of clusters for each year is highlighted using a consistent colour palette: green shades denote \(K=3\), and orange shades denote \(K=4\).
Early seasons display stronger support for \(K=3\), indicating a relatively fluid elite tier, while later years increasingly favour \(K=4\), reflecting the emergence of a more stratified competitive hierarchy during the Big Four era. More recent seasons seem to support once again the three block solution.
}
\label{tab:Kposterior_across_seasons}
\end{table}

\paragraph*{Early 2000s.}
Between 2000 and 2005, the three-block solution is consistently dominant (green cells in Tab.~\ref{tab:Kposterior_across_seasons}), suggesting a relatively fluid and competitive upper tier. During this phase, multiple players rotated through the top cluster from one season to the next, reflecting a more open contest among elite players.

\begin{figure}[htpb]
 \centering
 \includegraphics[width=\linewidth]{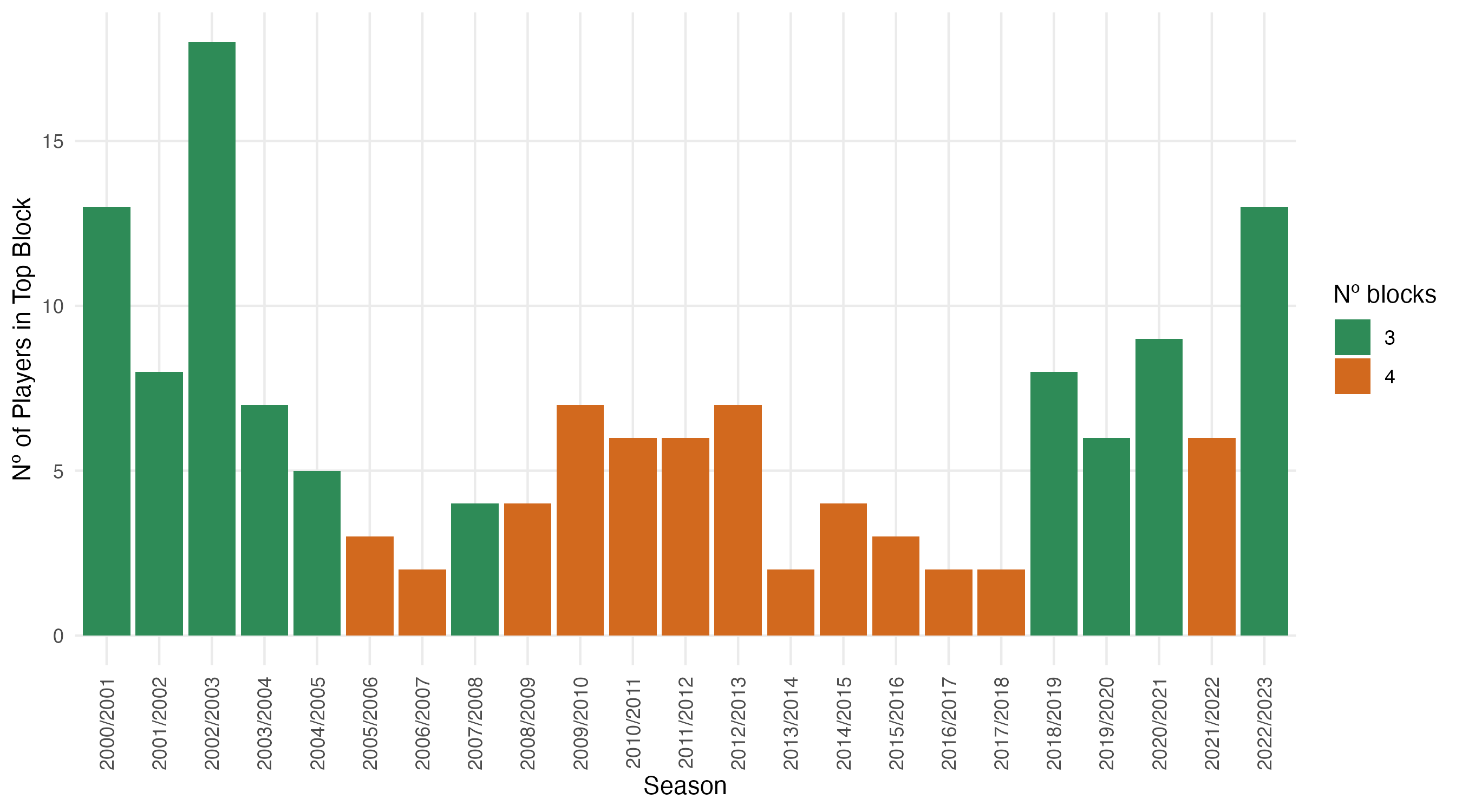}
\caption{
Estimated number of players assigned to the top cluster in each ATP season during the 2000/2001--2022/2023 ATP seasons.
The top block is substantially larger in the early 2000s, indicating a broader and more competitive elite group. 
From 2006 onward, the size of the top cluster shrinks markedly, coinciding with the rise and consolidation of the Big Four, who dominated the upper tier with little rotation. 
In 2018, the top block begins to expand again, signalling the end of that concentrated dominance and the re-emergence of a more open competition structure.
}

 \label{fig:n_players_top_block}
\end{figure}

Figure~\ref{fig:n_players_top_block} supports this reading by showing the size of the top block across seasons. In the early 2000s, the top cluster is noticeably larger, indicating a broader elite. This is a direct signal of competitiveness at the top: a more inclusive top tier implies less dominance by a fixed few.

\begin{figure}[htpb]
 \centering
 \includegraphics[width=\linewidth]{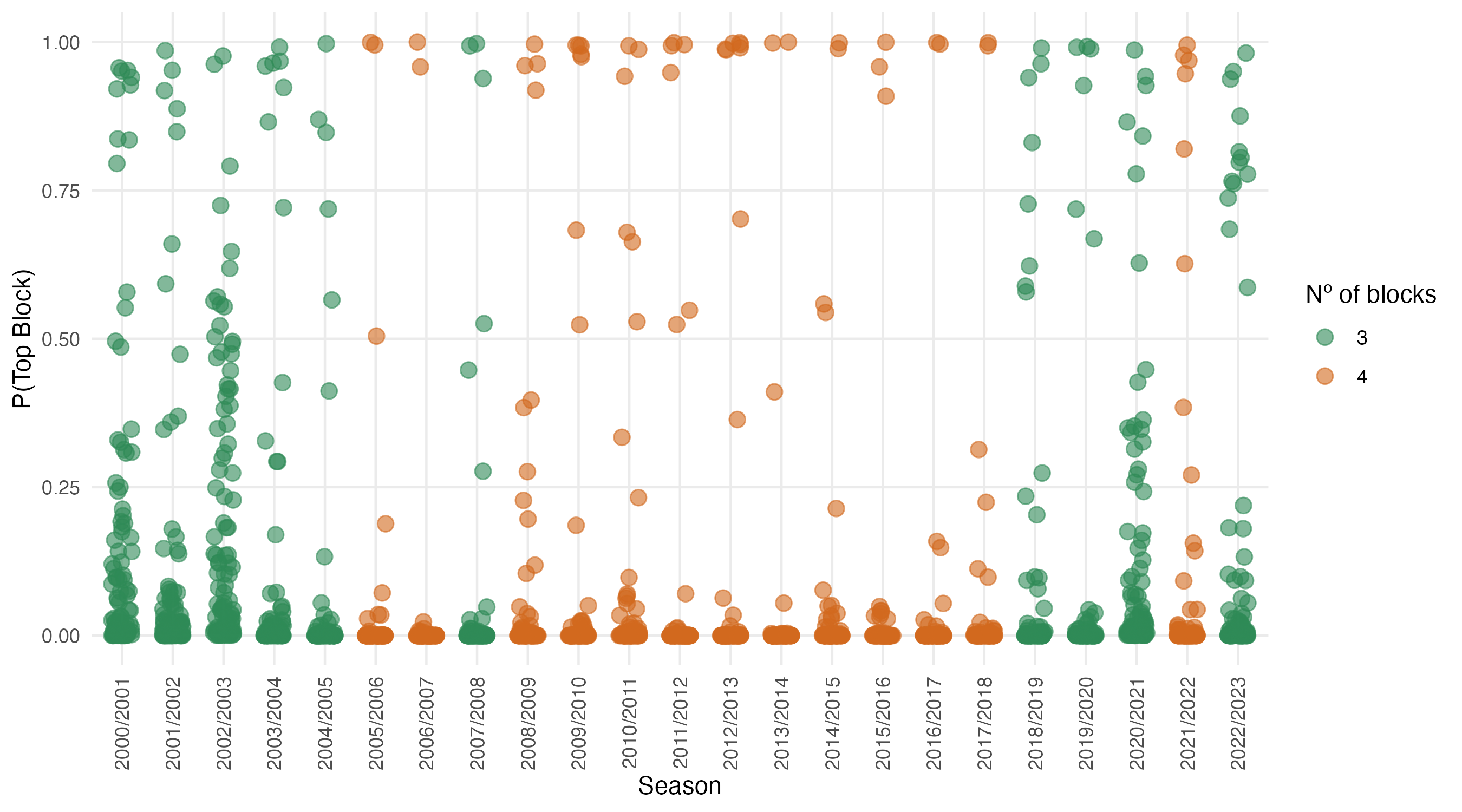}
\caption{
Posterior probability of belonging to the top cluster, by player and season, during the 2000/2001--2022/2023 ATP seasons. 
Each point represents a player's posterior inclusion probability in the top block for a given season; points are horizontally jittered to avoid overlap. 
Early seasons display a diffuse pattern with many players showing intermediate probabilities, consistent with a fluid and competitive elite. 
From the mid-2000s to 2018, only a small group---the Big Four---maintain probabilities close to one, reflecting near-constant dominance and reduced permeability of the top cluster. 
Around 2010, only a few points appear, offering limited evidence of a transition compared to the clearer regimes before and after. 
After 2020, intermediate probabilities become more frequent again, suggesting the re-emergence of a broader and more contested elite.
}\label{fig:p_top_across_years}
\end{figure}

\paragraph*{Mid 2000s to 2015.}
Starting in 2006, the model increasingly favours four-block solutions, and the top cluster size shrinks abruptly, marking the emergence and consolidation of the Big Four -- a tightly held elite with little rotation. The big four start to consistently dominate outcomes, and this is reflected both in the reduced top-cluster size and player-level posterior inclusion probabilities, which are examined in Figure~\ref{fig:p_top_across_years}. 

Here, each point represents the posterior probability that a given player belonged to the top block in a given season. During the early 2000s, top-cluster inclusion is more diffuse: several players exhibit intermediate probabilities, suggesting a more permeable elite. But in the following decade, a small number of players begin to dominate the top cluster with probabilities consistently near 1.

The sparsity of intermediate probabilities during this era reinforces this picture: most players were either clearly in or out of the top block, with little in-between. These fluctuations across both block size and individual inclusion probabilities offer a consistent and data-driven narrative of the Big Four era.

\paragraph*{From 2018 to recent years}
After 2018, the pattern loosens again: more players appear in the middle range of top-cluster inclusion probabilities, and the elite cluster begins to broaden. Taken together, Table~\ref{tab:Kposterior_across_seasons}, Figure~\ref{fig:n_players_top_block}, and Figure~\ref{fig:p_top_across_years} reinforce a coherent narrative. The Big Four era is characterized by both a structural contraction of the elite cluster and consequent subtle expansions that are non-trivial to detect or quantify. In the next section, we introduce a summary indicator designed to capture and formalize this dynamic.

\subsection{Measuring Competitive Balance}
\label{sec:competitive-balance}

To quantify changes in competitive balance over time, we measure the Shannon entropy
of the composition of the top block (block~1) at each MCMC iteration \(t\).
Let \(m_{1}^{(t)}\) denote the number of players assigned to block~1 at iteration~\(t\),
and \(m_{k}^{(t)}\) the number in block~\(k\), for \(k = 1, \ldots, K^{(t)}\).
We define the proportions
\(
p_k^{(t)} = \frac{m_k^{(t)}}{\sum_{\ell=1}^{K^{(t)}} m_\ell^{(t)}}\) for $k = 2, \ldots, K^{(t)}$. The entropy of the top-block composition is then
\begin{equation}
\label{eq:entropy_t}
H^{(t)} = - \sum_{k=1}^{K^{(t)}} p_k^{(t)} \log p_k^{(t)}.
\end{equation}
Low entropy (\(H^{(t)} \approx 0\)) indicates concentrated dominance by block~1,
whereas high entropy reflects a more even distribution of players across blocks. For interpretability across iterations with varying \(K^{(t)}\), we report
the normalized entropy
\(
\widetilde{H}^{(t)} = \frac{H^{(t)}}{\log K^{(t)}} \in [0,1],
\)
which equals~0 under complete dominance and~1 when cluster sizes are uniform.

\begin{figure}[htpb]
 \centering
 \includegraphics[width=0.95\linewidth]{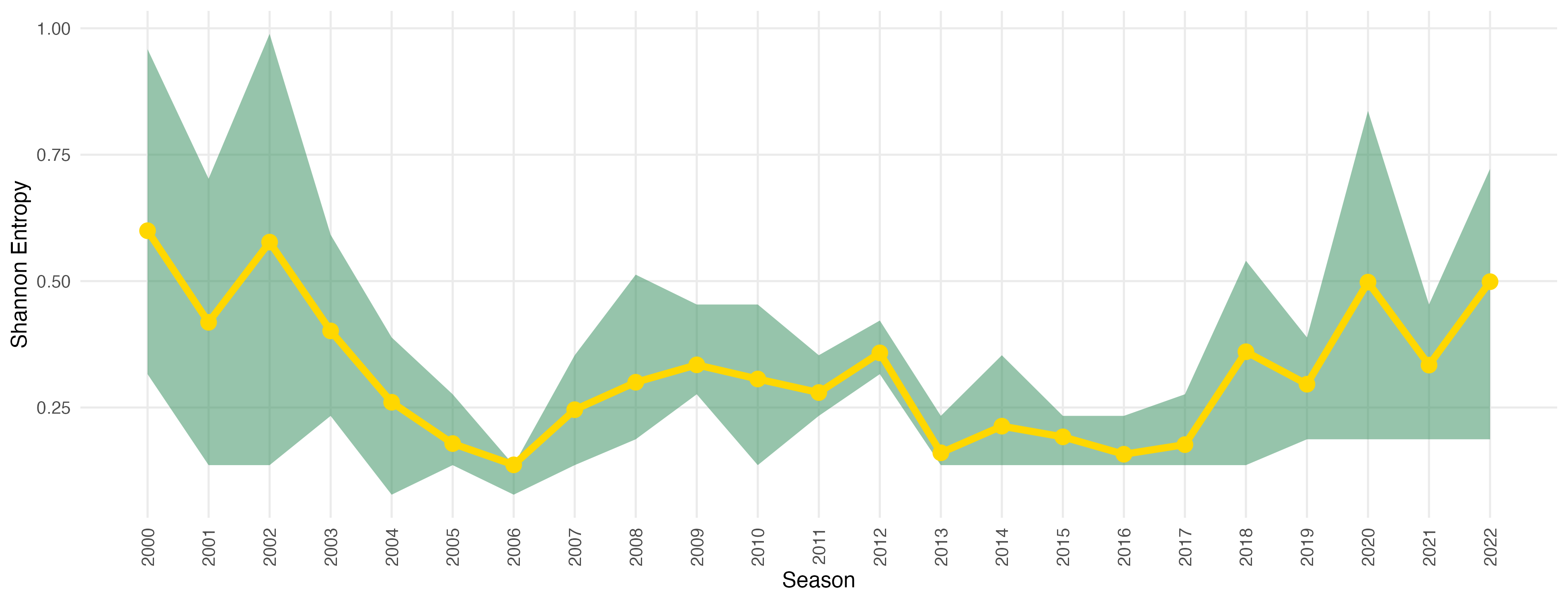}
\caption{
Posterior mean of the Shannon entropy of top-cluster composition across seasons, with 95\% credible intervals. 
Entropy quantifies the degree of competitive balance: high values indicate a more contested situation, with less predictable outcomes, and in turn, a more competitive sport; low values reflect concentrated dominance by few players. 
The temporal trajectory reveals three phases: high entropy in the early 2000s (a fluid competitive landscape), a sharp decline during the Big Four era (2006--2018) marking strong concentration at the top, and a post-2019 rebound indicating renewed balance and greater turnover among elite players.
}
 \label{fig:entropy_plot}
\end{figure}

Figure~\ref{fig:entropy_plot} displays the posterior mean normalized entropy over time, with 95\% credible intervals. The trajectory closely mirrors our earlier findings: competitive balance was highest in the early 2000s, collapsed during the Big Four era, and resurged after 2018. This metric offers a compact yet powerful diagnostic of competitive balance–and can be readily applied to other sports or domains to assess the degree of concentration or uncertainty in competitive hierarchies.

\section{Conclusion and Future Directions}
\label{sec:conclusion}

This work has introduced a Bayesian SBM for pairwise competitive outcomes, grounded in the Bradley--Terry framework and enriched by a nonparametric prior over partitions. At its core, the BT--SBM imposes that players in the same block have identical strength, i.e., $\lambda_i = \lambda_j \iff x_i = x_j$. This assumption induces a clustered ordering where co-clustered players share the same position in the ranking. This structure offers a principled and interpretable summary of competitive hierarchies by blending notions of similarity and dominance.

Our model was motivated by and applied to two decades of ATP men’s tennis. 
The analysis uncovered interpretable temporal trends -- most notably the emergence, consolidation, and eventual loosening of a dominant elite (the so-called Big Four). The results consistently supported a clustered ranking of players, where the top tiers showed stable membership and sharp separations, while the middle and lower tiers exhibited more diffuse and uncertain boundaries. The model’s ability to express uncertainty in both strength and cluster assignment represents a clear methodological advantage over classical approaches.

At the same time, the work acknowledges the limitations already highlighted in Section~\ref{sect:data}. Pairwise outcomes are sparse, sometimes inconsistent, and flatten away contextual information such as playing surface, player condition, or tournament setting. Moreover, the Bradley-Terry framework assumes linear stochastic transitivity, effectively reducing competition to a one-dimensional skill scale. Our formulation partially addresses these issues by introducing rank-indifference and latent clustering, but many dimensions of competitive heterogeneity remain unmodeled.

From a modelling standpoint, several extensions emerge as natural and valuable. First, our analysis treated each season independently. A richer formulation would incorporate \emph{temporal dependence} in both cluster allocations 
$\mathbf x$ and strength parameters $\bm \lambda$, enabling smooth evolution of player skill and cluster structure. Hidden Markov models, Gaussian processes, or autoregressive nonparametric priors (e.g., dynamic Dirichlet processes) are promising approaches.

Second, the current model captures heterogeneity entirely through latent structure. Incorporating \emph{covariates}--both at the edge level (e.g., surface type, tournament category, home advantage) and node level (e.g., age, handedness, ranking)--would address the present inability to account for observed factors that clearly shape outcomes.

Third, the assumption of linear stochastic transitivity remains restrictive. Competitive systems frequently display intransitivities or context dependence. Models that relax this assumption, for instance through weak or stochastic transitivity constraints \citep{lee2025pairwisecomparisonsstochastictransitivity, Spearing_2023}, offer a natural direction.

Fourth, the current model conditions on the observed match counts \(n_{ij}\), focusing solely on modelling  win outcomes \(w_{ij} \mid n_{ij}\). In professional tennis, however, the frequency of matches between players is itself an informative and dynamic process. High-ranked players tend to play more matches -- and against stronger opponents -- which creates a feedback loop between exposure and success that is not captured by the present framework. A more comprehensive model would treat \(n_{ij}\) as endogenous to the competitive process.

Finally, although our empirical focus was tennis, the BT--SBM framework is broadly applicable. It can be employed in any setting wherever pairwise directed comparisons are observed -- ranging from animal dominance in ecology \citep{baldassarre2023bradley}, to political preferences \citep{LOEWEN2012212}, transportation modelling  \citep{hatzinger2007transport}, and performance evaluation in AI systems \citep{ChiangChatbot2024}. Notably, this approach provides a natural diagnostic for \emph{competition uncertainty} across time or contexts, as shown by our use of entropy to quantify elite concentration.

In summary, the BT--SBM constitutes a first step toward a richer understanding of competitive structures. Its combination of interpretability, uncertainty quantification, and probabilistic coherence makes it a compelling tool for analysing systems of competition. At the same time, it opens several avenues for future work, in particular the incorporation of covariates, temporal dynamics, and more flexible transitivity assumptions.

\section*{Code and Reproducibility}
\label{sec:code}

\noindent
All code to reproduce the analyses, figures, and tables is openly available. We separate the
\emph{analysis workflow} and the \emph{general-purpose package} as follows:

\begin{itemize}
  \item \textbf{Reproducibility repository} (\texttt{BT--SBM-Bradley-Terry-Stochastic-Block-Model}):\\
  \href{https://github.com/laposanti/BT-SBM-Bradley-Terry-Stochastic-Block-Model}{github.com/laposanti/BT-SBM-Bradley-Terry-Stochastic-Block-Model}
  
  \item \textbf{R package} (\texttt{BTSBM}):\\
  \href{https://laposanti.github.io/BTSBM/}{laposanti.github.io/BTSBM/}
\end{itemize}

\begin{acks}[Acknowledgments]
For the purpose of Open Access, the author has applied a CC BY public copyright license to any Author Accepted Manuscript (AAM) version arising from this submission.
\end{acks}

\begin{funding}
This publication has emanated from research conducted with the financial support of Taighde Éireann -- Research Ireland under Grant number 18/CRT/6049. The Insight Centre for Data Analytics is supported by Science Foundation Ireland under Grant Number 12/RC/2289$\_$P2.
\end{funding}

\sdescription{This Supplement contains additional derivations, full conditional distributions for the Gibbs sampler, prior sensitivity analyses and hyperprior parameter settings, the derivation of the mean and variance of the Gnedin prior, details of the simulation study, supplementary figures referenced in the main text, and the code and R package used to reproduce all results.}

\bibliographystyle{imsart-nameyear} % Style BST file
\bibliography{Main.bib}       % Bibliography file (usually '*.bib')

\appendix
% No \appendix needed here; just start sections
\section{Derivations of full conditionals}
\label{sect:appendix_derivations}

In this appendix, we derive the full conditionals used in the Gibbs sampler for our augmented Bradley--Terry model with block structure. We exploit the latent variables \(Z_{ij} \sim \Gamma(n_{ij}, \lambda_{x_i} + \lambda_{x_j})\) as described in the augmentation scheme of Eq.~\eqref{eq:single_product_likelihood}. We aim to derive the full conditional distributions for the block strengths \(\lambda_k\) and the cluster assignments \(x_i\), based on the joint augmented likelihood:
\[
\mathcal{L}(\mathbf{W}, \mathbf{Z} \mid \bm{\lambda}, \mathbf{x},\mathbf{N}) \propto 
\prod_{i<j} \lambda_{x_i}^{w_{ij}} \lambda_{x_j}^{n_{ij} - w_{ij}} \cdot Z_{ij}^{n_{ij} - 1} \exp\left\{ -(\lambda_{x_i} + \lambda_{x_j}) Z_{ij} \right\}. \tag{A.1}
\]
Constants not involving \(\bm{\lambda}\), \(\mathbf{x}\), or \(\mathbf{Z}\) are omitted throughout.

\subsection{\texorpdfstring{Full conditionals for $\boldsymbol{\lambda}$}{Full conditionals for lambda}}
\label{sect:lambda_derivation}

We isolate the terms in A.1 involving \(\lambda_{x_i}\):
\[
\ell (\mathbf{W}, \mathbf{Z} \mid \bm{\lambda}, \mathbf{x},\mathbf{N}) 
\propto \underbrace{\sum_{i \ne j} w_{ij} \log \lambda_{x_i}}_{\text{(I)}} 
- \underbrace{\sum_{i < j} (\lambda_{x_i} + \lambda_{x_j}) Z_{ij}}_{\text{(II)}}.
\]

We rewrite each part in terms of individual-level quantities.

\subsubsection{Term (I).} 
Since \(\log \lambda_{x_i}\) is constant in \(j\), we factor it out:
\[
\sum_{i \ne j} w_{ij} \log \lambda_{x_i} = 
\sum_i \left( \sum_{j \ne i} w_{ij} \right) \log \lambda_{x_i} 
= \sum_i w_i \log \lambda_{x_i},
\]
where \(w_i = \sum_{j \ne i} w_{ij}\) is the total number of wins by individual \(i\).

\subsubsection{Term (II).} 
Each pair \((i,j)\) contributes \(\lambda_{x_i} Z_{ij} + \lambda_{x_j} Z_{ij}\), so the sum becomes:
\[
\sum_{i<j} (\lambda_{x_i} + \lambda_{x_j}) Z_{ij} 
= \sum_i \lambda_{x_i} \sum_{j \ne i} Z_{ij} 
= \sum_i \lambda_{x_i} Z_i,
\]
where \(Z_i = \sum_{j \ne i} Z_{ij}\).

\subsubsection{Final Form.} 
The complete-data log-likelihood, up to constants, becomes:
\[
\ell (\mathbf{W}, \mathbf{Z} \mid \bm{\lambda}, \mathbf{x}) \propto \sum_i \left[ w_i \log \lambda_{x_i} - \lambda_{x_i} Z_i \right].
\]

Now, recall the prior \(\lambda_k \sim \Gamma(a, b)\). Collecting the terms in the likelihood by block \(k\), let \(I_k = \{i : x_i = k\}\). The contribution of \(\lambda_k\) becomes:
\[
\ell(\lambda_k \mid \cdot) \propto 
\lambda_k^{\sum_{i \in I_k} w_i} 
\exp\left\{ -\lambda_k \sum_{i \in I_k} Z_i \right\}.
\]

Combining it with the Gamma prior:
\[
p(\lambda_k \mid \mathbf{W}, \bm{Z},\mathbf{x} ) \propto 
\lambda_k^{a - 1 + \sum_{i \in I_k} w_i} 
\exp\left\{ - \lambda_k (b + \sum_{i \in I_k} Z_i) \right\},
\]
we obtain as a posterior:
\[
\lambda_k \mid \mathbf{W}, \bm{Z},\mathbf{x}\sim 
\Gamma\left(a + \sum_{i \in I_k} w_i,\; b + \sum_{i \in I_k} Z_i\right).
\]

\subsection{Full Conditional for \texorpdfstring{$x_i$}{x}}
\label{sect:x_derivation}

We now derive the conditional probability of assigning player \(i\) to block \(k \in \{1, \dots, K\} \cup \{K+1\}\). Suppose \(i\) is temporarily removed from its current block.
\subsubsection{Existing Block \(k \in \{1, \dots, K\}\).}

Split the product in A.1 into terms involving \(i\) and those that do not:
\[
\mathcal{L}
= C(\mathbf{W})
\Biggl[\prod_{j \neq i} 
\underbrace{\Lambda_{ij}(k)}_{\text{depends on } k}
\Biggr]
\Biggl[\prod_{\substack{p<q \\ p,q \neq i}} \Lambda_{pq}\Biggr],
\]
where \(C(\mathbf{W})\) collects terms independent of \(\mathbf{x}\) and \(\boldsymbol{\lambda}\), and the second product is constant with respect to \(k\).  
For \(j > i\), one has
\[
\Lambda_{ij}(k)
= \lambda_k^{w_{ij}} 
  \lambda_{x_j}^{n_{ij}-w_{ij}} 
  Z_{ij}^{\,n_{ij}-1}
  e^{-\lambda_k Z_{ij}}
  e^{-\lambda_{x_j} Z_{ij}}.
\]
For \(j < i\), the roles of \(w_{ij}\) and \(n_{ij}-w_{ij}\) swap, but the exponent of \(\lambda_k\) still contributes either \(w_{ij}\) or \(n_{ij}-w_{ij}\).  
Summing over all \(j \neq i\) therefore yields
\[
\prod_{j \neq i} \Lambda_{ij}(k)
\propto
\lambda_k^{w_i}
\exp(-\lambda_k Z_i).
\]
\noindent
The prior assignment probability from the Gibbs-type prior is
\[
\Pr(x_i = k \mid \mathbf{x}_{-i})
\propto
\psi_{n,K}(m_{-i,k} - \sigma),
\]
where \(m_{-i,k}\) is the size of cluster \(k\) excluding node \(i\).  
Hence, the full conditional for \(x_i = k\) becomes
\[
p(x_i = k \mid \cdot)
\propto 
\psi_{n,K}(m_{-i,k} - \sigma)
\lambda_k^{w_i}
\exp(-\lambda_k Z_i).
\]

\subsubsection{New Block \(k = \{K + 1\}\).}

Here, \(x_i\) is assigned to a new cluster. We integrate $\lambda_k$ out:
\[
\int_0^\infty 
\lambda_k^{w_i} e^{-\lambda_k Z_i} \cdot \frac{b^a}{\Gamma(a)} \lambda_k^{a - 1} e^{-b \lambda_k} d\lambda_k
= \frac{b^a \Gamma(a + w_i)}{\Gamma(a)} (b + Z_i)^{-(a + w_i)} \quad \text{for}\; k = K+1.
\]

The prior for a new cluster is \(\psi_{n, K+1}\), so the full conditional becomes:
\[
p(x_i = K + 1 \mid \cdot) \propto 
\psi_{n,K+1} \cdot \frac{b^a \Gamma(a + w_i)}{\Gamma(a)} (b + Z_i)^{-(a + w_i)}.
\]

\noindent Together, these define the unnormalized probabilities for \(x_i \in \{1, \dots, K+1\}\), which are normalized to yield the full conditional distribution over block assignments.

\section{Scale alignment and prior standard deviation}
\label{app:role_prior_scale_clean}

In this appendix section, we provide a detailed motivation for the heuristic choice of fixing the Gamma prior’s rate as 
\(b=\exp\{\psi(a)\}\) and for choosing the shape parameter \(a\) in an interpretable way.
Block strengths are modelled as \(\lambda_k \sim \mathrm{Gamma}(a,b)\) 
-- using the shape–rate parametrization. 
Due to identifiability concerns (see Sect.~\ref{sect:identifiability}), the sampler renormalizes the occupied blocks at each iteration so that their log-strengths have arithmetic mean equal to zero, i.e.
\[
\frac{1}{K}\sum_{k=1}^K \log \lambda_k = 0 
\qquad \Longrightarrow \qquad
\mathbb{E}[\log \lambda_k \mid \mathbf{W},\mathbf{x}] = 0.
\]
Given this zero-induced expectation, it becomes natural to define the hyperprior so that the a-priori expected value of \(\log \lambda_k \) is zero as well. 

This moment-matching construction can be interpreted as a simple \emph{hyperprior heuristic}.
Rather than placing the prior on the raw block strengths \(\lambda_k\),
we work on \(\log \lambda_k\), and then we match the prior and posterior means:
\[
\mathbb{E}[\log \lambda_k \mid a, b, \mathbf{x}] 
\;=\; \mathbb{E}[\log \lambda_k \mid \mathbf{W}, a, b, \mathbf{x}]
\;=\; 0.
\]
From a modelling perspective, this alignment ensures that the prior and the likelihood operate on 
the same scale.  
From a computational standpoint, it prevents systematic up- or down-scaling in the creation of 
new clusters–an effect that would otherwise depend on arbitrary choices of \(a\) and \(b\).  

\smallskip
Formally, for $\lambda_k\sim \Gamma(a,b)$,
\[
\mathbb{E}[\log \lambda_k \mid a,b] = \psi(a) - \log b,
\]
where \(\psi(a)\) is the \emph{digamma function}, i.e.\ the derivative of 
\(\log \Gamma(a)\) \citep[see][]{joramsoch2025}.
By setting this expectation to zero, we define \(b_0\) as
\[
b_0 \coloneqq \exp\{\psi(a)\},
\]
thereby ensuring that the prior mean of \(\log \lambda_k\) is consistent with the global scale imposed by the sampler.

\smallskip
To study the role of this choice in the propensity to create new clusters (see Eq.~\eqref{eq:new_cluster_prob}),
define the bias \(\delta = \psi(a) - \log b\),
the discrepancy between the a-priori expectation and the normalization-induced (a-posteriori) expectation, which is zero. 
We want to see how the likelihood contribution to the creation of a new cluster (Eq.~\eqref{eq:new_cluster_prob}) changes with respect to $\delta$. Therefore, we compute
\begin{align*}
    \Delta_i(\delta) &\coloneqq \log L_i (K+1 \mid a,b) - \log L_i (K+1 \mid a,b_0) \\
  & -a\,\delta 
  - (a+w_i)\!\left[
      \log\!\big(e^{\psi(a)-\delta}+Z_i\big) - 
      \log\!\big(e^{\psi(a)}+Z_i\big)
    \right] \quad \text{where } b_0 = \exp{\psi(a)}.
\end{align*}
Differentiating gives
\[
\Delta_i'(\delta) \;=\; -a \;+\; (a+w_i)\frac{b}{\,b+Z_i\,}, 
\qquad b \;=\; e^{\psi(a)-\delta},
\]
so that \(-a < \Delta_i'(\delta) < w_i\).
Here \(w_i = \sum_{j\neq i} w_{ij}\) denotes the total number of successes for item \(i\), and
\(Z_i\) is the aggregate augmentation mass attached to \(i\).
Specifically, with \(n_{ij}\) matches between \(i\) and \(j\) and the Gamma augmentation, we have
\[
Z_{ij}\,\big|\,n_{ij},\lambda_{x_i},\lambda_{x_j} \;\sim\; \Gamma\!\left(n_{ij},\,\lambda_{x_i}+\lambda_{x_j}\right),
\]
as defined in Eq.~\eqref{eq:z_update}. We report it for convenience also:
\[
Z_i \;=\; \sum_{j\neq i} Z_{ij},
\qquad 
\mathbb{E}\!\left[Z_i \,\middle|\, \bm \lambda, \mathbf{x} \right] 
\;=\; \sum_{j\neq i} \frac{n_{ij}}{\lambda_{x_i}+\lambda_{x_j}}.
\]
Thus \(Z_i\) scales with the total number of matches played by item \(i\) (larger when \(i\) has faced many opponents)
and is down-weighted when \(i\) or its opponents have large \(\lambda\)-values.
Intuitively, \(Z_i\) can be interpreted as an ``strength-adjusted'' measure of the sample size associated with \(i\): more matches and weaker opponents yield a larger \(Z_i\).

\begin{figure}[htpb]
\centering
\includegraphics[width=\textwidth]{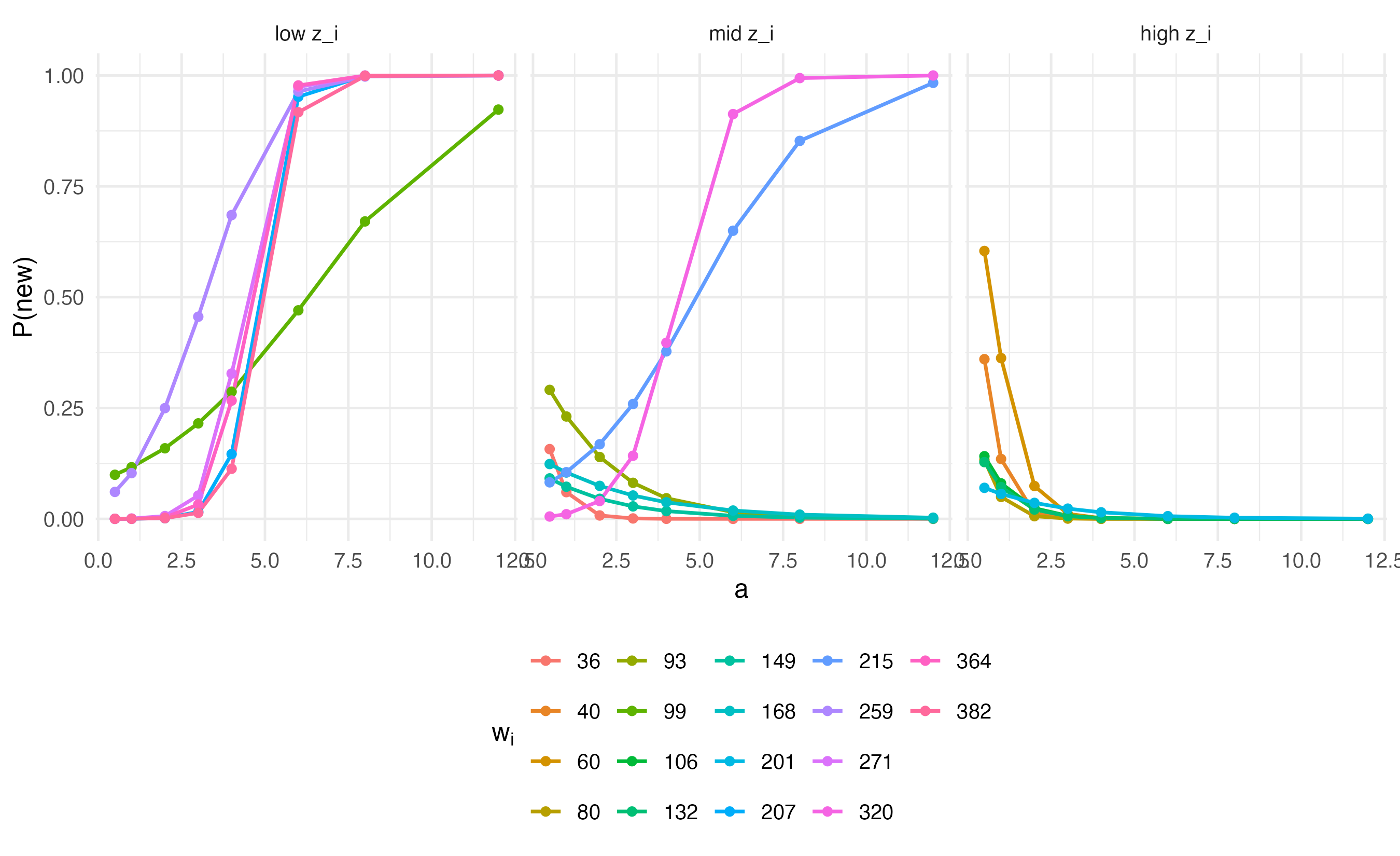}
\caption{How the probability of creating a new cluster varies with $a$ in the misaligned case 
($b=1$, $\delta=\psi(a)$). 
The three panels correspond to empirical tertiles of $Z_i$: 
low ($Z_i \le \widehat q_{1/3}$), mid ($\widehat q_{1/3} < Z_i \le \widehat q_{2/3}$), 
and high ($Z_i > \widehat q_{2/3}$), 
where $\widehat q_{1/3}$ and $\widehat q_{2/3}$ are the sample 33\% and 66\% quantiles of 
$\{Z_i\}_{i=1}^n$. 
For small $Z_i$, the evidence for a new cluster is systematically inflated (left panel), 
whereas for large $Z_i$ it is deflated (right panel). 
Hence, setting $b=1$ in the prior for $\log\lambda$ induces a non-linear, $Z$-dependent effect 
on the probability of creating a new block.
}
\label{fig:pnew_lines_b1}
\end{figure}

\begin{figure}[htpb]
\centering
\includegraphics[width=\textwidth]{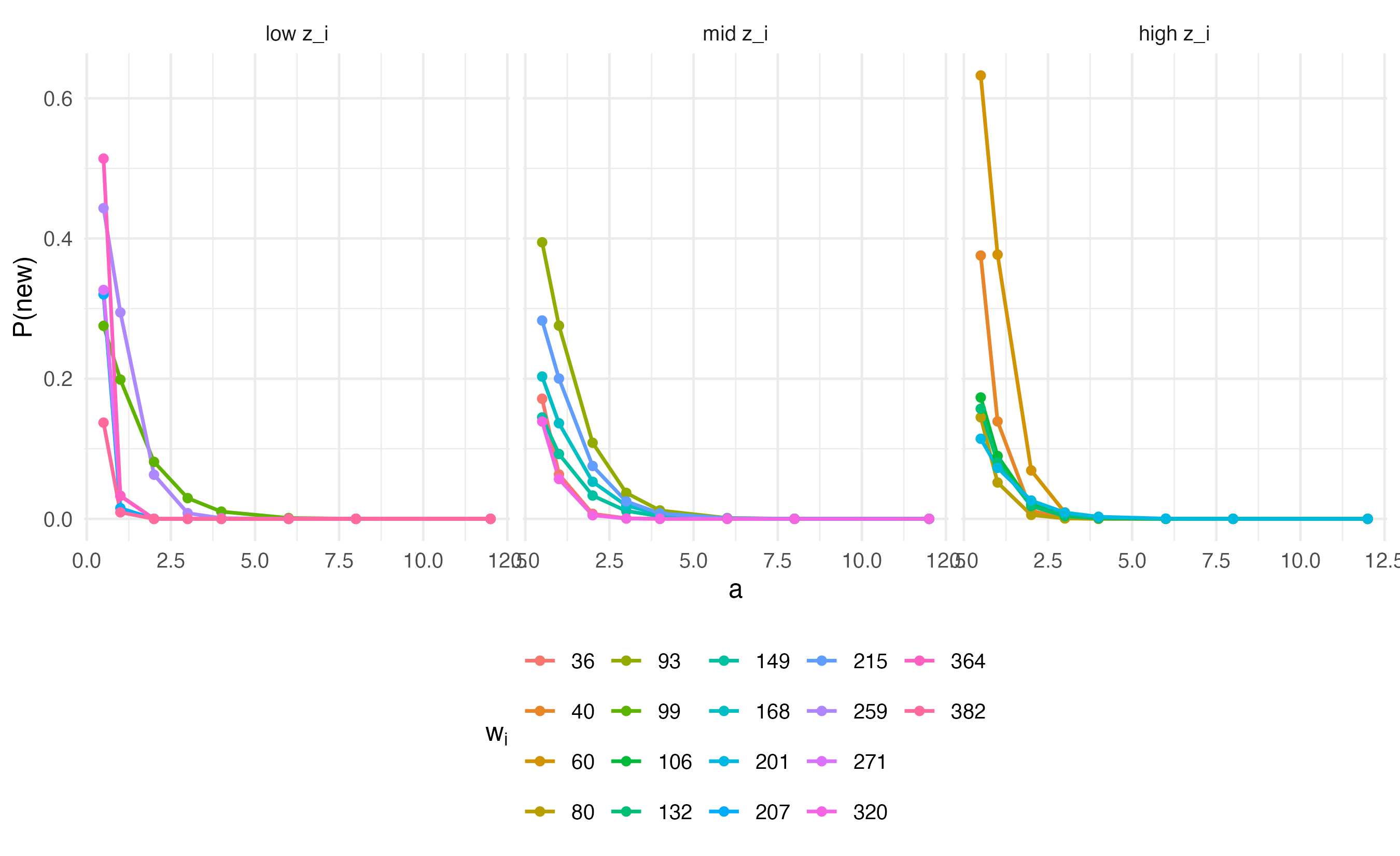}
\caption{How the probability of creating a new cluster varies with $a$ in the aligned case 
($b_0=\exp\{\psi(a)\}$, $\delta=0$). 
The three panels correspond to empirical tertiles of $Z_i$: 
low ($Z_i \le \widehat q_{1/3}$), mid ($\widehat q_{1/3} < Z_i \le \widehat q_{2/3}$), 
and high ($Z_i > \widehat q_{2/3}$), 
where $\widehat q_{1/3}$ and $\widehat q_{2/3}$ are the sample 33\% and 66\% quantiles of 
$\{Z_i\}_{i=1}^n$. 
In all panels, the prior alignment removes the $Z$-dependent bias in the probability of creating 
a new cluster, producing consistent behaviour across $Z$ values.
}
\label{fig:pnew_lines_bpsia}
\end{figure}

\smallskip
The sign of \(\Delta_i'(\delta)\) reveals two distinct regimes, clearly visible in 
Figures~\ref{fig:pnew_lines_b1}--\ref{fig:pnew_lines_bpsia}.
When \(Z_i\) is small, 
\(\tfrac{b}{b+Z_i}\approx 1\) and \(\Delta_i'(\delta)\approx w_i>0\):
increasing \(\delta\) (i.e.\ choosing \(b<e^{\psi(a)}\)) \emph{inflates} the odds of creating a new cluster, 
making such items more likely to initiate new blocks. 
This effect appears in Figure~\ref{fig:pnew_lines_b1} as the upward shift of the curves for 
low-\(Z_i\) items under the misaligned prior (in this specific case \(b=1\), but other $b$ values would yield the same effect).
When \(Z_i\) is large, 
\(\tfrac{b}{b+Z_i}\approx 0\) and \(\Delta_i'(\delta)\approx -a<0\):
increasing \(\delta\) then \emph{deflates} the new-cluster odds, 
as seen in the lower placement of the high-\(Z_i\) curves in the same figure.  
Hence, any \(\delta\neq 0\) induces a \(Z_i\)-dependent tilt in the new-cluster likelihood, 
while the aligned choice \(\delta=0\) (Figure~\ref{fig:pnew_lines_bpsia}) removes this effect entirely, 
causing all curves to stabilize along the same monotone and downward sloping trajectory.

\smallskip

\subsection{\texorpdfstring{Prior standard deviation on log--$\boldsymbol{\lambda}$}%
{Prior standard deviation on log--lambda}}

\smallskip
Once the mean of \(\log\lambda_k\) is fixed, the remaining degree of freedom is its 
\emph{standard deviation} of \(\log \lambda_k\) which, for \(\lambda_k\sim\mathrm{Gamma}(a,b)\), corresponds to:
\[
\mathrm{Var}[\log \lambda_k] = \psi_1(a), \qquad
\tau \equiv \mathrm{SD}(\log \lambda_k) = \sqrt{\psi_1(a)},
\]
where \(\psi_1(a)\) is the \emph{trigamma function}, the derivative of the digamma 
\citep[][Eq.~5.15.1]{joramsoch2025}. 
Hence, the shape parameter \(a\) controls the prior heterogeneity on \(\log\lambda_k\): 
large \(a\) (small \(\tau\)) yields concentrated priors with nearly homogeneous block strengths, 
while small \(a\) (large \(\tau\)) allows for more heterogeneous blocks.

We therefore treat $\tau$ as an intuitive hyperparameter controlling the prior heterogeneity of block strengths. The mapping between $a$ and $\tau$ follows from inverting the following expression:
\[
\psi_1(a) = \tau^2 
\qquad \Longrightarrow \qquad 
a = \psi_1^{-1}(\tau^2).
\]
Representative values are shown in Table~\ref{tab:a_tau}.

\begin{table}[htpb]
\centering
\begin{tabular}{ccc}
\toprule
$a$ & $\tau = \mathrm{SD}(\log \lambda_k)$ & $\mathrm{SD}(\lambda_k) = \sqrt{a}/b$ \\
\midrule
\cellcolor{gray!10}{6} & \cellcolor{gray!10}{0.43} & \cellcolor{gray!10}{0.44}\\
5 & 0.47 & 0.50\\
\cellcolor{gray!10}{4} & \cellcolor{gray!10}{0.53} & \cellcolor{gray!10}{0.57}\\
3 & 0.63 & 0.69\\
\cellcolor{gray!10}{2} & \cellcolor{gray!10}{0.80} & \cellcolor{gray!10}{0.93}\\
1 & 1.28 & 1.78\\
\bottomrule
\end{tabular}\caption{Relationship between the Gamma shape parameter $a$ and the corresponding standard deviation of 
$\log \lambda_k$, denoted by $\tau = \sqrt{\psi_1(a)}$. 
The second column reports $\tau$, which measures prior variability on the log scale. 
The rightmost column shows the standard deviation of $\lambda_k$ under the scale alignment 
$b = \exp\{\psi(a)\}$, illustrating how dispersion on the log scale translates into 
actual variability in the block strengths.
}
\label{tab:a_tau}
\end{table}

In practice, moderate values such as $\tau \in [0.5,0.8]$ (i.e.\ $a \in [4,2]$) provide a good balance between flexibility and stability.
From Table~\ref{tab:a_tau}, these correspond to a standard deviation of $\mathrm{SD}(\lambda_k)\approx0.6-0.9$ under the aligned prior $b=\exp{\psi(a)}$, meaning that block strengths typically vary within roughly a 50\%–100\% range around their geometric mean. Thus, $\tau$ acts as a single interpretable control knob for the prior heterogeneity of $\lambda_k$, while the alignment $b=\exp{\psi(a)}$ ensures this variability remains invariant to the scale of $\lambda_k$, identifiable, and well behaved with respect to the auxiliary varible $Z$ .

\clearpage

\section{Closed-form mean and variance of \texorpdfstring{$K$}{K} under the Gnedin model}\label{sect:mean_var_K_gnedin}

In this appendix, we report the closed-form expressions for the mean and the variance of $K$ under the Gnedin prior, where $K$ denotes the \emph{sample} number of clusters (that is, the number of groups represented in the observed data). The calculations below are a routine specialization of standard manipulations for Gibbs–type partitions \citep{favaro2013}. 
We work in the finite–type case (\(\sigma=-1\)) known as the Gnedin model, for which the number of clusters has the explicit pmf
\[
p(K=k\mid n,\gamma)
= \binom{n}{k}\,
\frac{(1-\gamma)_{k-1}\,(\gamma)_{\,n-k}}{(1+\gamma)_{\,n-1}},
\qquad k=1,\dots,n,
\]
as reported in Eq.~\eqref{eq:gnedin_prior}, with \((a)_m=\Gamma(a+m)/\Gamma(a)\) the rising Pochhammer symbol; see \citet[Sec.~6, Eq.~(9)]{gnedin2010}. 
As in the general Gibbs–type literature, we combine (i) index–shift identities for binomials, 
\(k\binom{n}{k}=n\binom{n-1}{k-1}\) and \(k(k-1)\binom{n}{k}=n(n-1)\binom{n-2}{k-2}\), 
with (ii) expressing expectations in factorial form, and (iii) repeated applications of the 
Chu–Vandermonde (binomial–Pochhammer) identity
\[
\sum_{r=0}^{N}\binom{N}{r}(a)_r(b)_{N-r}=(a+b)_N,
\]
exactly in the manner used in the BNP literature to collapse partition sums; see 
\citet[App.~A.2, Lemma~A.1]{lijoi2008a} for a precise statement and its use in Gibbs–type derivations, 
and \citet{favaro2013} for related generalizations and the use of steps (i)--(iii). 
Nothing essentially new is given in this \(\sigma=-1\) setting: the same general off–the–shelf identities in \citep{favaro2013} lead directly to the closed–form mean and variance reported below, and we include the explicit formulas only for completeness and ease of reference.

\subsection{First factorial moment and mean}

Using the identity \(k\binom{n}{k}=n\binom{n-1}{k-1}\), we write: 
\begin{align*}
\mathbb{E}[K]
&=\sum_{k=1}^{n} k\,p(K=k\mid n,\gamma)
=\frac{n}{(1+\gamma)_{n-1}}
  \sum_{k=1}^{n}\binom{n-1}{k-1}(1-\gamma)_{k-1}(\gamma)_{(n-1)-(k-1)}.
\end{align*}
Setting \(r=k-1\) and applying the Chu–Vandermonde identity with 
\(a=1-\gamma\), \(b=\gamma\), and \(N=n-1\), we obtain
\[
\mathbb{E}[K]
=\frac{n}{(1+\gamma)_{n-1}}\,(1)_{n-1}
= \frac{\Gamma(n+1)\,\Gamma(1+\gamma)}{\Gamma(n+\gamma)}.
\]
This expression coincides with the mean used for prior specification in recent 
applications of the Gnedin model \citep[e.g.][]{Legramanti_2022}, 
and can be viewed as the \(\sigma=-1\) specialization of the general Gibbs–type expectation derived 
via factorial–moment expansions in \citet{favaro2013}.

\subsection{Second factorial moment}

Proceeding analogously and using \(k(k-1)\binom{n}{k}=n(n-1)\binom{n-2}{k-2}\), 
we can derive the second factorial moment as: 
\[
\begin{aligned}
\mathbb{E}[K(K-1)]
&=\frac{n(n-1)}{(1+\gamma)_{n-1}}
\sum_{k=2}^{n}\binom{n-2}{k-2}(1-\gamma)_{k-1}(\gamma)_{n-k}.
\end{aligned}
\]
Letting \(r=k-2\) and factoring out one term from the rising factorial,  
\((1-\gamma)_{k-1}=(1-\gamma)(2-\gamma)_{k-2}\), we obtain
\[
\mathbb{E}[K(K-1)]
=\frac{n(n-1)(1-\gamma)}{(1+\gamma)_{n-1}}
\sum_{r=0}^{n-2}\binom{n-2}{r}(2-\gamma)_{r}(\gamma)_{(n-2)-r}.
\]
Applying the Chu–Vandermonde identity again, now with \(a=2-\gamma\), 
\(b=\gamma\), and \(N=n-2\), yields
\[
\mathbb{E}[K(K-1)]
=\frac{n(n-1)(1-\gamma)\,(n-1)!}{(1+\gamma)_{n-1}}.
\]
This calculation mirrors the repeated use of the Vandermonde–Pochhammer convolution 
adopted by \citet{lijoi2008a} to simplify factorial–moment 
sums in Gibbs–type models, here specialized to the finite–type case \(\sigma=-1\).

\subsection*{Variance}

Ordinary and factorial moments are related through  
\(\mathbb{E}[K^2]=\mathbb{E}[K(K-1)]+\mathbb{E}[K]\) 
\citep[see, e.g.,][]{pitman2006, favaro2013},  
so the variance follows immediately:
\[
\mathrm{Var}(K)
=
\frac{n(n-1)(1-\gamma)\,(n-1)!}{(1+\gamma)_{n-1}}
+\frac{\Gamma(n+1)\Gamma(1+\gamma)}{\Gamma(n+\gamma)}
-\left(\frac{\Gamma(n+1)\Gamma(1+\gamma)}{\Gamma(n+\gamma)}\right)^2.
\]

\subsubsection{Specification for the present context.}
For the empirical setting considered here, with $n=105$ and $\gamma=0.8$, we have
\[
\mathbb{E}[K]\;=\;\frac{\Gamma(106)\Gamma(1.8)}{\Gamma(105.8)}\;\approx\;2.36,
\qquad
\mathrm{Var}(K)\;\approx\;45.95.
\]
Thus, while the prior mean is about two to three clusters, the variance is quite large, 
yielding a heavy–tailed distribution that allows for additional clusters when supported by the data. 
This behaviour reflects the role of $\gamma$ in the partition prior: smaller values of $\gamma$ 
inflate the variability of $K$, favouring richer partitions, whereas larger values concentrate the prior 
on more parsimonious configurations \citep[see also][]{gnedin2010, favaro2013, DeBlasi2015}.

\section{Simulation Study}
\label{sec:simulation_appendix}

To validate the proposed model, we conduct a simulation study designed to test whether the BT--SBM can recover both the true number of blocks and the latent partition under a correctly specified data-generating process. The experiment mirrors the empirical ATP setting while allowing full control over the underlying cluster structure.

\subsection{Data Generation}
\label{sec:sim_data_generation}

We generate synthetic datasets to assess the model's ability to recover the true number of clusters \(K^\ast\) and partition \(\mathbf{x}^\ast\). 
The number of items is fixed to \(n = 150\), and the target number of clusters varies as \(K^\ast \in \{3, \ldots, 10\}\).
For each target \(K^\ast\), we generate a structured but interpretable dataset according to the following steps. 

First, we simulate the edge structure to reproduce the sparsity observed in the real data: for each unordered pair \((i,j)\), an edge is included with probability \(0.5\); otherwise, the pair is excluded. This ensures that roughly half of the potential matches are realized, producing a network density comparable to that of the empirical ATP season.

Second, we induce clusters of approximately equal size by repeating the true labels sequentially,
\(
x_i^\ast = (1,2,\dots,K^\ast,1,2,\dots,K^\ast,\ldots)
\)
up to \(n = 150\). This yields equally-sized clusters allowing us to isolate the model’s clustering accuracy from block-imbalance effects.

Third, we assign block strengths deterministically as equally spaced values 
\(\lambda_k^\ast \in [0.1, 3]\), \(k = 1,\dots,K^\ast\), ensuring clear yet realistic differences between cluster-level abilities.

Finally, conditional on the generated cluster structure, we sample the number of matches between connected pairs from a 
\(\mathrm{Poisson}(5)\) distribution, and simulate match outcomes \(w_{ij}\) under the Bradley–Terry- SBM likelihood. 

\begin{algorithm}[htbp]
\caption{Synthetic data generation for the BT--SBM. 
Edges match real-data sparsity; clusters are balanced; 
block strengths are equally spaced for clear separation.}
\label{alg:generating1}
\begin{algorithmic}[1]
\STATE \textbf{Require} $n = 105$, target $K^\ast \ge 2$
\FORALL{pairs $(i,j)$ with $1 \le i < j \le n$}
  \STATE Draw $e_{ij} \sim \mathrm{Bernoulli}(0.5)$
  \IF{$e_{ij}=1$}
    \STATE Draw $n_{ij} \sim \mathrm{Poisson}(5)$
  \ELSE
    \STATE $n_{ij} \gets 0$
  \ENDIF
\ENDFOR
\STATE Assign cluster labels $x_i^\ast = (1,2,\dots,K^\ast,1,2,\dots)$ truncated at $n=105$
\STATE Set $\lambda_k^\ast \gets 0.1 + (k-1)\,\dfrac{3 - 0.1}{K^\ast - 1}$ for $k=1,\dots,K^\ast$
\FORALL{pairs $(i,j)$ with $n_{ij}>0$}
  \STATE Draw $w_{ij} \sim \mathrm{Binomial}\!\left(n_{ij},\;
  \dfrac{\lambda_{x_i^\ast}}{\lambda_{x_i^\ast} + \lambda_{x_j^\ast}}\right)$
\ENDFOR
\end{algorithmic}
\end{algorithm}

\subsection{Recovery of the true number of clusters}
\label{sec:K_recovery}

For each $K^\ast \in \{3,\dots,10\}$, we generate 23 replicates and fit the model using 30{,}000 MCMC iterations with \(\gamma = 4\).  
The posterior distribution of \(K\) is summarized by its posterior mode \(\widehat{K}\), as reported in Table~\ref{tab:clustering_performance}.  

\begin{table}[htpb]
\centering
\caption{Cross--tabulation of the estimated number of clusters \(\widehat{K}\) (rows) versus the true number \(K^\ast\) (columns). Diagonal entries represent exact recovery; off--diagonal entries correspond to near misses.}
\label{tab:clustering_performance}
\begin{tabular}{lrrrrrrrr}
\toprule
\(\widehat{K} \,\backslash\, K^\ast\) & 3 & 4 & 5 & 6 & 7 & 8 & 9 & 10 \\
\midrule
3 & 23 & - & - & - & - & - & - & -\\
4 & - & 23 & - & - & - & - & - & -\\
5 & - & - & 23 & - & - & - & - & -\\
6 & - & - & - & 23 & - & - & - & -\\
7 & - & - & - & - & 23 & - & - & -\\
8 & - & - & - & - & - & 22 & 1 & -\\
9 & - & - & - & - & - & 4 & 18 & 1\\
10 & - & - & - & - & - & 1 & 13 & 9\\
\bottomrule
\end{tabular}
\end{table}

The results show that for \(K^\ast \le 9\), the model almost always recovers the correct number of clusters.  
For \(K^\ast = 10\), the posterior tends to slightly underestimate \(K\), typically returning \(\widehat{K} = 9\).  
This mild shrinkage is expected: as the number of blocks grows, the effective sample size per block decreases,  
leading to occasional merging of clusters during posterior inference.  
Nevertheless, in all scenarios, the true \(K^\ast\) remains within the 95\% posterior credible interval,  
indicating good posterior calibration.

\subsection{Recovery of the true partition}
\label{sec:partition_recovery}

We next assess how well the model recovers the true partition \(\mathbf{x}^\ast\).  
Posterior samples \(\mathbf{x}^{(t)}\) are relabelled to address label-switching using Algorithm~\ref{alg:label_switch},  
and summarized by the VI loss estimator, as reported in Sect.~\ref{sect:point_estimates}. Clustering accuracy is evaluated via the Adjusted Rand Index (ARI) \citep{Hubert_1985}, which ranges from 0 (no agreement) to 1 (perfect recovery).  

\begin{figure}[htpb]
  \centering
  \includegraphics[width=0.65\linewidth]{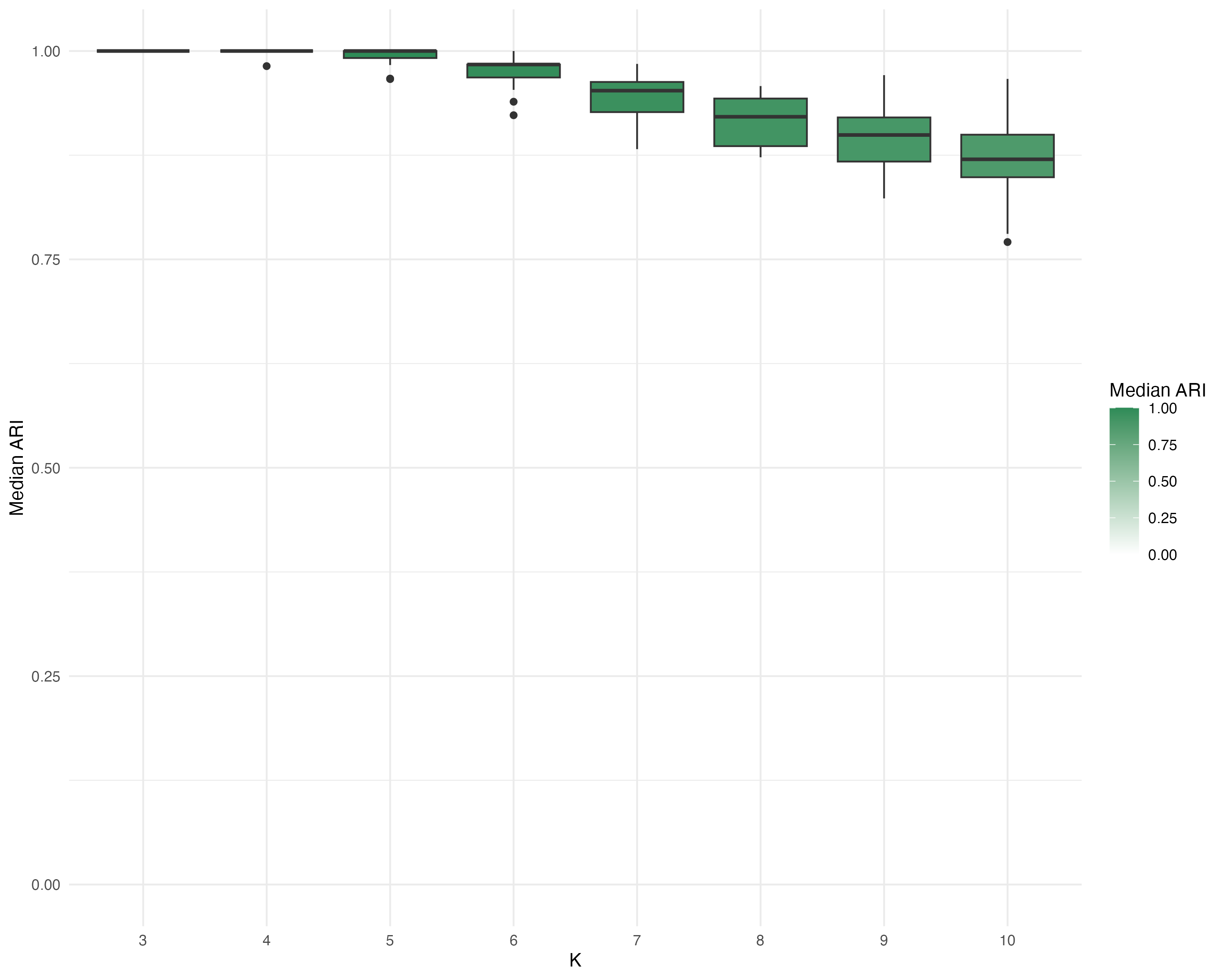}
  \caption{Adjusted Rand Index (ARI) across 23 replicates for each true number of clusters \(K^\ast\). 
  Each boxplot shows the distribution of ARI scores for a fixed \(K^\ast\).}
  \label{fig:ari_distribution}
\end{figure}

Figure~\ref{fig:ari_distribution} shows consistently high recovery accuracy.  
For \(K^\ast \le 9\), median ARI values exceed 0.9, with narrow dispersion across replicates.  
For \(K^\ast = 10\), performance slightly decreases (median ARI \(\approx 0.85\)), mainly due to the smaller sample size per block.  
Even in these cases, the estimated partitions remain close to the ground truth, and the VI loss estimator provides smooth, stable summaries of posterior clustering uncertainty.

\smallskip

Overall, the simulation study confirms that the BT--SBM equipped with a Gnedin prior reliably recovers both the number of clusters and the underlying partition structure when the model is correctly specified.

\section{Boundary partitions on the credible ball surface}
\label{appendix:edge-partitions}

\begin{figure}[htpb]
  \centering
  \includegraphics[width=\linewidth]{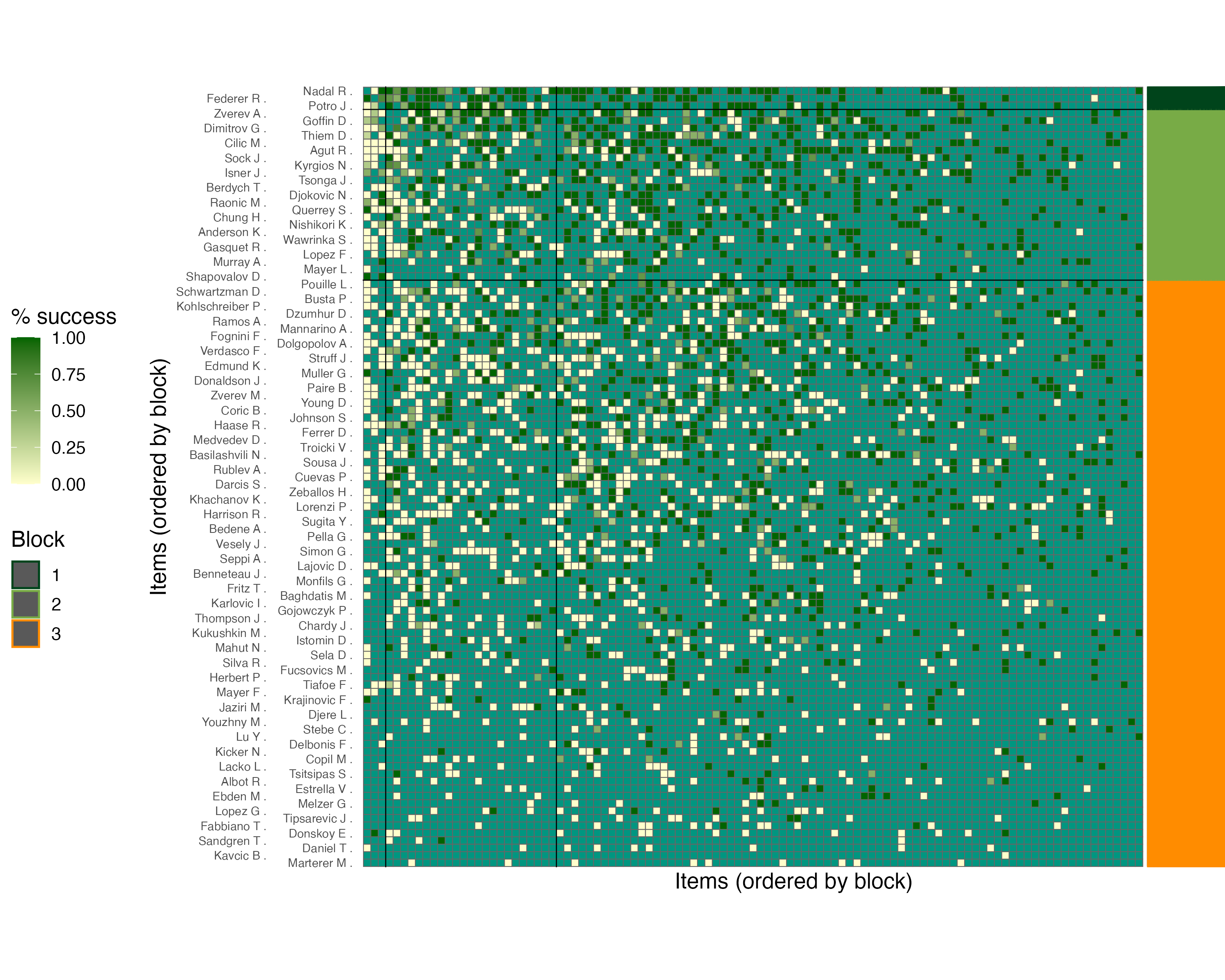}
  \caption{Adjacency matrix of the 2017/2018 season reordered according to the vertical upper bound (\(K=11\)) of the 95\% credible ball around \(\widehat{\mathbf{x}}\). This is the coarsest partition within the credible ball that is maximally distant from \(\widehat{\mathbf{x}}\). The only difference with respect to Fig.~\ref{fig:adjacency-reordered}, is the fact that the top block now includes also \emph{del Potro}, while other blocks remain identical. For more details about how to read this figure, check the caption of Fig.~\ref{fig:adjacency-reordered}.
  }
  \label{fig:v-ub}
\end{figure}

\begin{figure}[htpb]
  \centering
  \includegraphics[width=\linewidth]{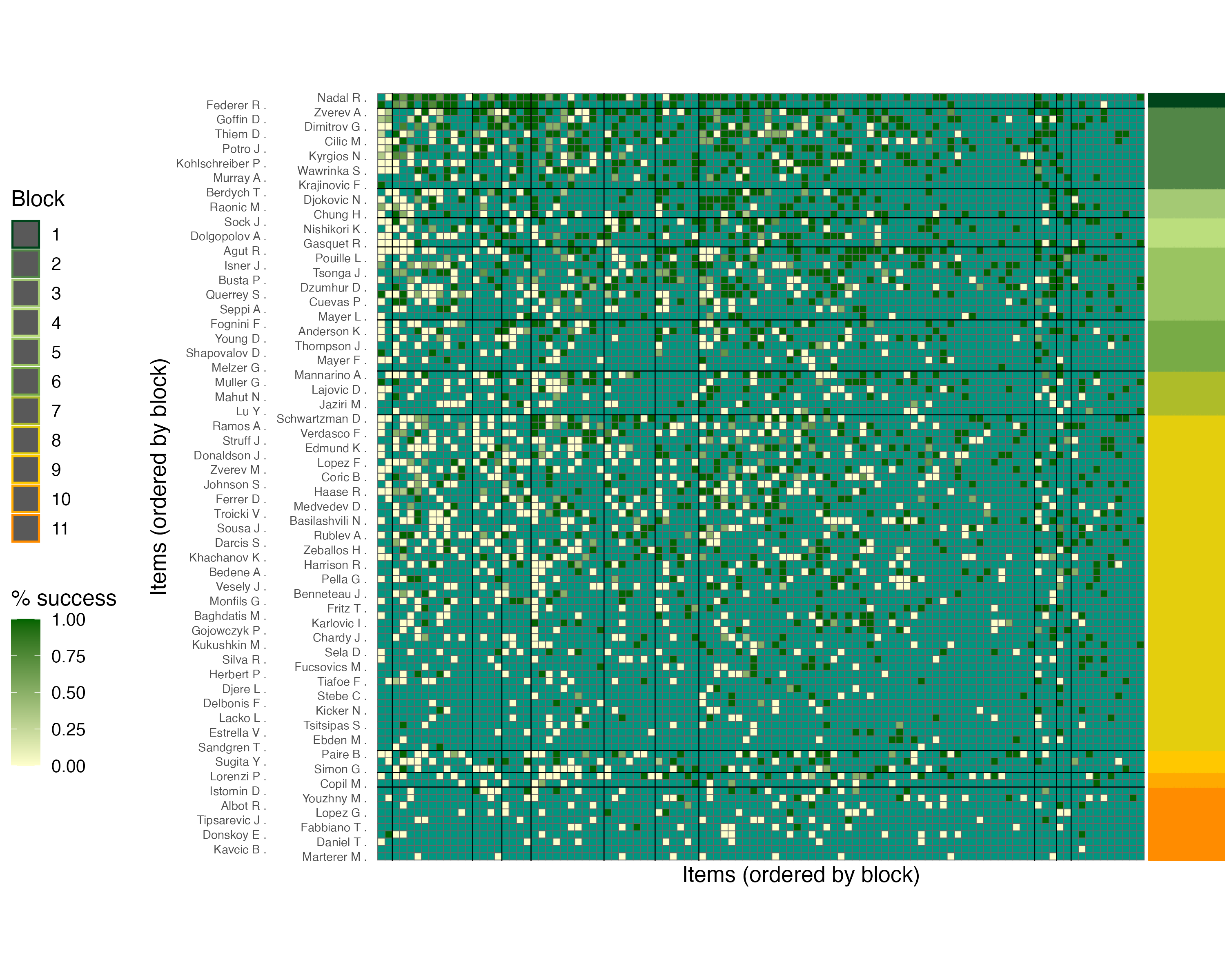}
  \caption{Adjacency matrix of the 2017/2018 season reordered according to the vertical lower bound (\(K=11\)) of the 95\% credible ball around \(\widehat{\mathbf{x}}\).  
This is the finest partition within the credible ball, where both mid-tier players and weaker players are split into several smaller groups, capturing more subtle distinctions in playing strength.  
The elite block remains narrow–still dominated by \emph{Nadal} and \emph{Federer}–while the weakest players continue to cluster at the bottom into a large block.  
The main change is a dense fragmentation of the middle, where the large mid-tier group in the point estimate divides into six sub-blocks, and a further differentiation at the lower end, where additional blocks emerge to represent varying strength levels among the weaker players.  
For more details on how to read this figure, see the caption of Fig.~\ref{fig:adjacency-reordered}.
}
  \label{fig:v-lb}
\end{figure}

\begin{figure}[htpb]
  \centering
  \includegraphics[width=\linewidth]{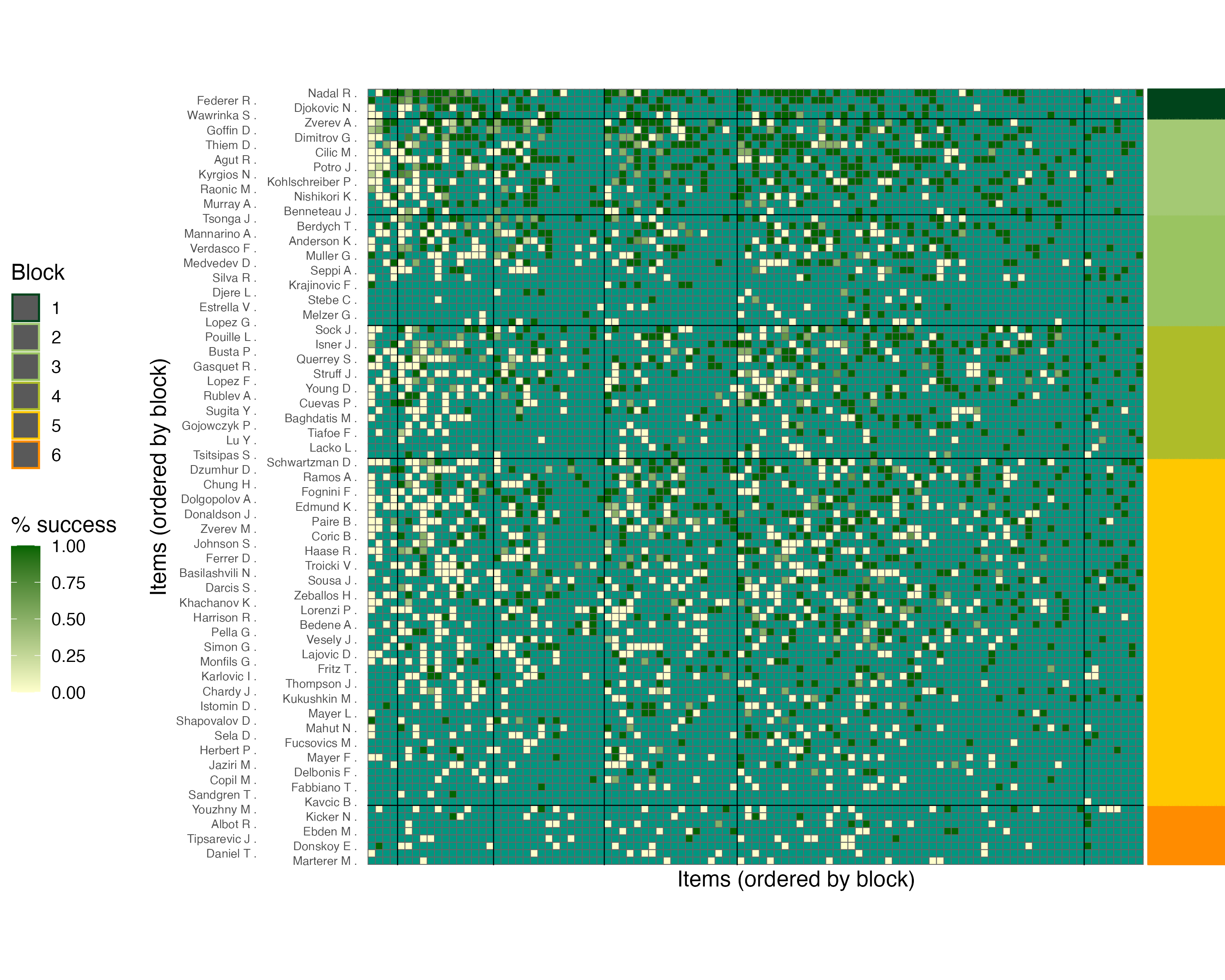}
  \caption{Adjacency matrix of the 2017/2018 season reordered according to the horizontal bound (\(K=6\)) of the 95\% credible ball around \(\widehat{\mathbf{x}}\).  
This configuration doubles the number of clusters relative to the VI estimate, producing a more even subdivision into blocks.  
The elite block slightly expands to include \emph{Djokovic} and \emph{Wawrinka}, while the overall cluster sizes become more balanced.  
As in the other configurations, the main differences arise at the boundaries between blocks, where most of the posterior uncertainty concentrates.  
For more details on how to read this figure, see the caption of Fig.~\ref{fig:adjacency-reordered}.
}
  \label{fig:horiz}
\end{figure}

To provide a clearer view of posterior uncertainty around the estimated partition \(\widehat{\mathbf{x}}\), we report the magnified components of Figure~\ref{fig:placeholder}, that is, the surface partitions of the 95\% credible ball defined in Section~\ref{sect:point_estimates}. 
Following \citet{Wade_2018}, these \emph{boundary partitions} correspond to clusterings that are maximally distant from \(\widehat{\mathbf{x}}\) under the VI loss, while still lying within the credible region. The distance employed here is the VI distance, which we omit to repeat for brevity. We display the three bounds reported in Figure~\ref{fig:placeholder}: 
\begin{enumerate}
\item \emph{Vertical upper bound}, reported in Fig.~\ref{fig:v-ub}, with its \(K=3\) blocks, is the most distant partition with the least amount of blocks (what we call the \emph{coarsest partition}) within the 95\% credible ball;
\item \emph{Vertical lower bound}, reported in Fig.~\ref{fig:v-lb}, showcases \(K=11\) blocks and represents the most distant partition with the maximum number of blocks (the \emph{finest partition}) within the credible region;
\item \emph{Horizontal bound}, reported in Fig.~\ref{fig:horiz} with \(K=6\), is the most distant partition within the credible ball irrespective of \(K\).
\end{enumerate}
Together they characterize and narrow down the posterior uncertainty distribution with respect to the inferred partition $\widehat{\mathbf{x}}$

\subsection{Description and interpretation}

We begin by outlining what the three configurations share, and then describe where they differ.  
Across all of them, the elite block is sharply defined: \emph{Nadal} and \emph{Federer} appear in every configuration, while \emph{Djokovic}, \emph{Wawrinka}, and occasionally \emph{del Potro} join them depending on the resolution.  
These players also display the highest assignment uncertainty in Fig.~\ref{fig:ass-prob}, indicating that the model hesitates only over the precise composition of the elite, not its existence.  
At the opposite end of the hierarchy, the weakest block is always cohesive and clearly separated.  
Between these two extremes lies a consistent group of contenders striving to enter the elite.  
In sum, the basic \(K = 3\) tiered structure–elite, contenders, and weaker players–remains visible across all partitions, which mainly differ in how finely these three blocks are subdivided.

Beyond this variability in \emph{resolution}, the main differences among boundary partitions occur near the interfaces between the three principal blocks.  
Uncertainty concentrates at these transition regions – between the elite and contenders, and between the mid-tier and the weakest group–while the cores of each block remain stable.

For example, the \emph{horizontal bound} (Fig.~\ref{fig:horiz}, \(K = 6\)) introduces additional blocks precisely at these interfaces: parts of the upper mid-tier and of the weakest group detach to form two small intermediate clusters.  
In the \emph{vertical lower bound} (Fig.~\ref{fig:v-lb}, \(K = 11\)), this pattern becomes more pronounced.  
The weakest block in the point estimate splits into several subgroups–two minor and one more substantial–capturing finer distinctions among the least successful players.  
Yet even in these high-resolution configurations, the same three main regions are still identifiable, merely refined by additional layers at their borders.

\subsection{Summary and take-away}

The boundary partitions confirm a stable \emph{tiered structure} of player strengths, while revealing uncertainty in how finely those tiers should be divided.  
As the resolution increases, new blocks appear only at the boundaries, not in the core of the existing ones.  
This supports our modelling choice of \emph{tiered ranks}: the tiers themselves capture the robust features of the posterior, while fine-grained allocations across neighbouring blocks reflect quasi-arbitrary choices of resolution.  
In short, the model identifies who belongs to which broad level of the hierarchy–the tiered order is clear–but the exact placement of the boundaries remains uncertain, and that uncertainty is an integral part of the data-generating process.

\end{document}